\PassOptionsToPackage{prologue,dvipsnames}{xcolor}
\documentclass[acmsmall]{acmart}

\AtBeginDocument{%
  }

\setcopyright{cc}
\setcctype{by}
\acmDOI{10.1145/3729343}
\acmYear{2025}
\acmJournal{PACMSE}
\acmVolume{2}
\acmNumber{FSE}
\acmArticle{FSE073}
\acmMonth{7}
\received{2024-09-11}
\received[accepted]{2025-04-01}

\usepackage[T1]{fontenc}    
\usepackage{url}            
\usepackage{booktabs}       
\usepackage{amsfonts}       
\usepackage{nicefrac}       
\usepackage{microtype}      
\usepackage{multirow}
\usepackage{enumitem}
\usepackage{amsmath,amsfonts}
\usepackage{bm}
\usepackage{graphicx}
\usepackage{caption}
\usepackage{listings}
\usepackage{textcomp}
\usepackage{colortbl}
\usepackage{subcaption}
\usepackage{tcolorbox}
\usepackage{xspace}
\usepackage{flushend}
\usepackage{pifont}
\usepackage{wrapfig}
\usepackage{mdframed}
\usepackage{paracol}
\usepackage{needspace}
\usepackage{hyperref} 
\usepackage{threeparttable}
\usepackage[ruled,linesnumbered]{algorithm2e}
\usepackage{multicol}
\usepackage[dvipsnames]{xcolor}

\makeatletter
\DeclareRobustCommand\onedot{\futurelet\@let@token\@onedot}
\def\@onedot{\ifx\@let@token.\else.\null\fi\xspace}
\def\ie{{i.e}\onedot} 

\def\etal{{et al}\onedot}
\makeatother

\newcommand{\midsepremove}{\aboverulesep = 0mm \belowrulesep = 0mm}
    \midsepremove
    \newcommand{\midsepdefault}{\aboverulesep = 0mm \belowrulesep = 0mm}
    \midsepdefault

\newcommand{\responseref}[0]{\color{black}}
\newcommand{\responseline}[1]{\textcolor{black}{#1}}

\definecolor{darkBlue}{rgb}{0.000000,0.000000,0.545098}
\definecolor{darkGreen}{rgb}{0.000000,0.392157,0.000000}
\definecolor{DarkGray}{gray}{0.4}
\definecolor{javared}{rgb}{0.6,0,0} 
\definecolor{javagreen}{rgb}{0.25,0.5,0.35} 
\definecolor{javapurple}{rgb}{0.5,0,0.35} 
\definecolor{javadocblue}{rgb}{0.25,0.35,0.75} 
\definecolor{lightgray}{gray}{0.7}

\definecolor{shadecolor}{RGB}{150,150,150}
\definecolor{blueA}{RGB}{204,229,255}
\definecolor{redA}{RGB}{112,0, 0}
\definecolor{RED}{RGB}{255,0, 0}
\definecolor{lightred}{HTML}{FFEBEE}
\definecolor{lightblue}{HTML}{d3e9f7}

\definecolor{red1}{HTML}{c4483f}
\definecolor{red2}{HTML}{d56964}
\definecolor{red3}{HTML}{e48989}
\definecolor{red4}{HTML}{efa9ad}
\definecolor{red5}{HTML}{f8cacf}

\newcounter{finding}
\newenvironment{finding}
{
    \refstepcounter{finding}
	\begin{mdframed}[
    	nobreak=true,
    	linecolor=black,
    	roundcorner=12pt,
    	backgroundcolor=gray!05,
    	linewidth=0.3pt,
    	leftmargin=0.5em,
    	rightmargin=0.5em,
    	topline=true,
    	bottomline=true,
    	frametitlerule=true,
    	frametitlebackgroundcolor=gray!30,
    	frametitlerulecolor=gray,
    	frametitle=Finding \arabic{finding},
    	frametitleaboveskip=0.3em,
    	frametitlebelowskip=0.35em,
    	skipabove=5pt
	]
   \em
}
{
    \end{mdframed}
    \vspace{1em}
}

\newcommand{\smalltitle}[1]{\vspace{1mm}{\noindent\bf \textit{#1}}}

\usepackage{array}
\newcommand{\PreserveBackslash}[1]{\let\temp=\\#1\let\\=\temp}
\newcolumntype{C}[1]{>{\PreserveBackslash\centering}p{#1}}
\newcolumntype{R}[1]{>{\PreserveBackslash\raggedleft}p{#1}}
\newcolumntype{L}[1]{>{\PreserveBackslash\raggedright}p{#1}}

\newcommand{\rqone}{How do VLA models perform in popular robotic manipulation tasks?}

\newcommand{\RQone}{How Do VLA Models Perform in Popular Robotic Manipulation Tasks?}

\newcommand{\rqtwo}{How does the number of \responseline{confounding objects} affect a VLA model's performance?}

\newcommand{\RQtwo}{How Does the Number of \responseline{Confounding Objects} Affect a VLA Model's Performance?}

\newcommand{\rqthree}{Does the change in lighting conditions affect a VLA model's performance?}

\newcommand{\RQthree}{Does the Change in Lighting Conditions Affect a VLA Model's Performance?}

\newcommand{\rqfour}{Does the change of camera pose affect a VLA model's performance?}

\newcommand{\RQfour}{Does the Change of Camera Pose Affect a VLA Model's Performance?}

\newcommand{\rqfive}{How robust do VLA models perform against unseen objects?}

\newcommand{\RQfive}{How Robust Do VLA Models Perform Against Unseen Objects?}

\newcommand{\rqsix}{How robust do VLA models perform against task instruction mutations?}

\newcommand{\RQsix}{How Robust Do VLA Models Perform Against Task Instruction Mutations?}

\newcommand{\method}{\textsc{VLATest}}
\newcommand{\tool}{\method}

\definecolor{amii_night}{HTML}{003f58}
\definecolor{amii_sky}{HTML}{6c98ab}
\definecolor{amii_mustard}{HTML}{faa53c}

\begin{document}

\title{VLATest: Testing and Evaluating Vision-Language-Action Models for Robotic Manipulation}

\author{Zhijie Wang}
\orcid{0000-0003-4559-5426}
\affiliation{%
  \institution{University of Alberta}
  \city{Edmonton}
  \country{Canada}
}
\email{zhijie.wang@ualberta.ca}

\author{Zhehua Zhou}
\orcid{0000-0001-9542-4858}
\affiliation{%
  \institution{University of Alberta}
  \city{Edmonton}
  \country{Canada}
}
\email{zhehua1@ualberta.ca}

\author{Jiayang Song}
\orcid{0009-0008-7093-9781}
\affiliation{%
  \institution{University of Alberta}
  \city{Edmonton}
  \country{Canada}
}
\email{jiayan13@ualberta.ca}

\author{Yuheng Huang}
\orcid{0000-0003-3666-4020}
\affiliation{%
  \institution{The University of Tokyo}
  \city{Tokyo}
  \country{Japan}
}
\email{yuhenghuang42@g.ecc.u-tokyo.ac.jp}

\author{Zhan Shu}
\orcid{0000-0002-5933-254X}
\affiliation{%
  \institution{University of Alberta}
  \city{Edmonton}
  \country{Canada}
}
\email{zshu1@ualberta.ca}

\author{Lei Ma}
\orcid{0000-0002-8621-2420}
\affiliation{%
  \institution{The University of Tokyo}
  \city{Tokyo}
  \country{Japan}
}
\affiliation{%
  \institution{University of Alberta}
  \city{Edmonton}
  \country{Canada}
}
\email{ma.lei@acm.org}


\begin{abstract}

The rapid advancement of generative AI and multi-modal foundation models has shown significant potential in advancing robotic manipulation. Vision-language-action (VLA) models, in particular, have emerged as a promising approach for visuomotor control by leveraging large-scale vision-language data and robot demonstrations. However, current VLA models are typically evaluated using a limited set of hand-crafted scenes, leaving their general performance and robustness in diverse scenarios largely unexplored. To address this gap, we present {\method}, a fuzzing framework designed to generate robotic manipulation scenes for testing VLA models. Based on {\method}, we conducted an empirical study to assess the performance of seven representative VLA models. Our study results revealed that current VLA models lack the robustness necessary for practical deployment. Additionally, we investigated the impact of various factors, including the number of confounding objects, lighting conditions, camera poses, unseen objects, and task instruction mutations, on the VLA model's performance. Our findings highlight the limitations of existing VLA models, emphasizing the need for further research to develop reliable and trustworthy VLA applications.

\end{abstract}



\maketitle

\section{Introduction}







Robotic manipulation is widely regarded as one of the most important areas within cyber-physical systems (CPS). Over the past few decades, robotic manipulation and its applications have been implemented across various domains, such as industrial automation~\cite{dzedzickis2021advanced, goel2020robotics}, healthcare~\cite{kyrarini2021survey, holland2021service}, and logistics and warehousing~\cite{shamout2022conceptual, dhaliwal2020rise}. 
With the rapid advancement of AI techniques, researchers and practitioners have been exploring the integration of AI for planning~\cite{guo2023recent, rayhan2023artificial} and control~\cite{soori2023artificial, wakchaure2023application} in robotic manipulation. 
For instance, recent studies have employed deep reinforcement learning for robotics control to enhance the adaptability of the system against unseen scenarios~\cite{han2023survey, orr2023multi}. 

Meanwhile, the emergence of foundation models, such as large language models (LLMs) and vision-language models (VLMs), has introduced new opportunities for advancing AI-enabled robotic manipulation~\cite{singh2023progprompt, ding2023task}. 
In particular, LLMs and VLMs have demonstrated promising capabilities to participate in the loop of robotics system development and operation to enhance corresponding reasoning, perception, and task-planning abilities~\cite{huang2023instruct2act, liang2023code, driess2023palm}.
In contrast to the general purpose-oriented LLMs and VLMs, vision-language-action (VLA) models were developed exclusively to generate robot actions for manipulation based on visual observations from cameras and task instructions provided by users’ natural language input. Techniques such as reinforcement learning from human feedback (RLHF)~\cite{bai2022training} further enhance these models’ ability to interpret human intent from natural language commands. 
As a result, it is now possible to let robotics perform manipulation tasks simply by prompting the VLA models. 
Additionally, the large-scale pre-training nature of such foundation models makes it possible to adapt to a diverse range of downstream manipulation tasks without task- and environment-specific niche design. 
A recent study demonstrates that the state-of-the-art (SOTA) VLA model, RT-2~\cite{brohan2023rt}, can be extended to complex tasks such as symbol understanding and human recognition that were unseen during training~\cite{brohan2023rt}. 

However, the data-driven nature of VLA models makes them difficult to interpret, which hinders the safety and applicability of VLA models to more robotics applications in practice.
To enhance the reliability and trustworthiness of VLA models for robotic manipulation, the development of quality assurance techniques such as testing, debugging, and repairing has become an urgent need in both industrial and academic communities. 
In particular, comprehensive and adequate testing is essential for assessing a VLA model’s capabilities and limitations. 
Unfortunately, current VLA models are typically evaluated using only a limited set of hand-crafted scenes, where the comprehensiveness and effectiveness of such scenes are often inadequate, which subsequently leaves the general performance and behavior characteristics of VLA models largely unexplored.
Moreover, unlike conventional LLMs, which only process natural language text inputs, VLA models are empowered by the multimodality perception ability; therefore, when deploying VLA models in real-world scenarios, visual factors such as \responseline{confounding objects}, lighting conditions, and camera angles may significantly impact their performance. 
Despite this, there is currently a lack of deep understanding regarding the robustness of VLA models against such environmental factors.

\responseline{
To address this gap, we conducted a large-scale empirical study to evaluate the overall performance and robustness of popular VLA models.
We first propose {\method}, one of the earliest testing frameworks specifically designed for evaluating VLA models in robotic manipulation. }
We introduce a set of ten testing operators and design a generation-based fuzzing framework that automatically produces testing scenes to assess the performance and robustness of VLA models in specific robotic manipulation tasks. 
{\method} is implemented within the Maniskill2 simulation environment~\cite{gu2023maniskill2}. 
\responseline{
By leveraging {\method}, we aim to efficiently identify potential bugs in using VLA models for robotic manipulation, enabling comprehensive evaluation of VLA models under varying conditions.
}
We designed a set of research questions to examine key factors in reliable robotic manipulation with VLA: \textit{basic performance}, \textit{task complexity}, \textit{perception robustness}, \textit{OOD (out-of-distribution) robustness}, \responseline{and language model robustness}. 
Specifically, we investigated the following six research questions: 
\begin{itemize}[leftmargin=0.45in]
    \setlength\itemsep{0.5mm}
    \item[RQ1] \rqone
    \item[RQ2] \rqtwo
    \item[RQ3] \rqthree
    \item[RQ4] \rqfour
    \item[RQ5] \rqfive
    {\responseref{} \item[RQ6] \rqsix}
\end{itemize}
To investigate these research questions, we generated 18,604 testing scenes across four different robotic manipulation tasks. 
We selected \textit{seven} popular publicly available VLA models: RT-1-1k, RT-58k, RT-400k~\cite{brohan2022rt}, RT-1-X~\cite{padalkar2023open}, Octo-small, Octo-base~\cite{team2024octo}, and OpenVLA-7b~\cite{kim2024openvla}.
The experiments included a total of 78,604 rounds of simulation execution, which took more than 580 GPU hours. 

Our analysis shows that current VLA models exhibit \textbf{subpar performance} across the four robotic manipulation tasks studied. As the number of \responseline{confounding objects} increases, it becomes more challenging for these models to accurately locate and manipulate the correct objects. Additionally, we observed that the subject VLA models lack robustness when subjected to changes in lighting conditions and camera angles, performing much worse compared to the default settings. Meanwhile, we found that VLA models with large-scale pre-training generally show better robustness against such changes. Finally, we found that these VLA models struggle significantly with unseen objects, exhibiting a significant drop in performance compared to manipulating seen objects. \responseline{Current VLA models also often fell short in understanding the same task instruction paraphrased in different ways.} Our findings provide practical guidelines for developers working with VLA models on specific robotic manipulation tasks and underscore the need for more advanced VLA models, as well as novel quality assurance techniques to enhance the robustness and reliability of the VLA models. 

In summary, this paper makes the following contributions:

\begin{itemize}[leftmargin=*]
    \setlength\itemsep{0.5mm}

    {\responseref{}

    \item \textbf{An empirical study.} We conducted a large-scale empirical study to evaluate the performance and robustness of seven popular VLA models across four robotic manipulation tasks under various conditions, including variations in the number of \responseline{confounding objects}, lighting conditions, camera poses, and the manipulation of both seen and unseen objects.

    \item \textbf{A testing framework.} To conduct our empirical study, we designed and implemented a generation-based fuzzing framework, {\method}, to test VLA models by incorporating various operators in robotic manipulation tasks.
    }

    \item \textbf{Implications.} We discussed the challenges and limitations of current VLA models, along with the implications and future opportunities for enhancing their robustness and reliability.

    \item \textbf{Artifacts.} Our artifacts, including the replication packages and the generated testing scenes, are available on a GitHub repository: \href{https://github.com/ma-labo/VLATest}{https://github.com/ma-labo/VLATest}.
    
\end{itemize}

\section{Background: Vision-Language-Action Models}




Vision-Language-Action (VLA) models~\cite{brohan2022rt, padalkar2023open, team2024octo, kim2024openvla} are a type of deep neural network that takes natural language input from the user as task instructions and visual input from a camera as observations. The output of a VLA model is a set of actions to achieve the designated manipulation according to the task instructions, such as moving a robotic arm’s joints and opening the gripper.

\subsection{VLA Model for Robotic Manipulation}

\begin{figure}[t]
    \centering
    \includegraphics[width=\linewidth]{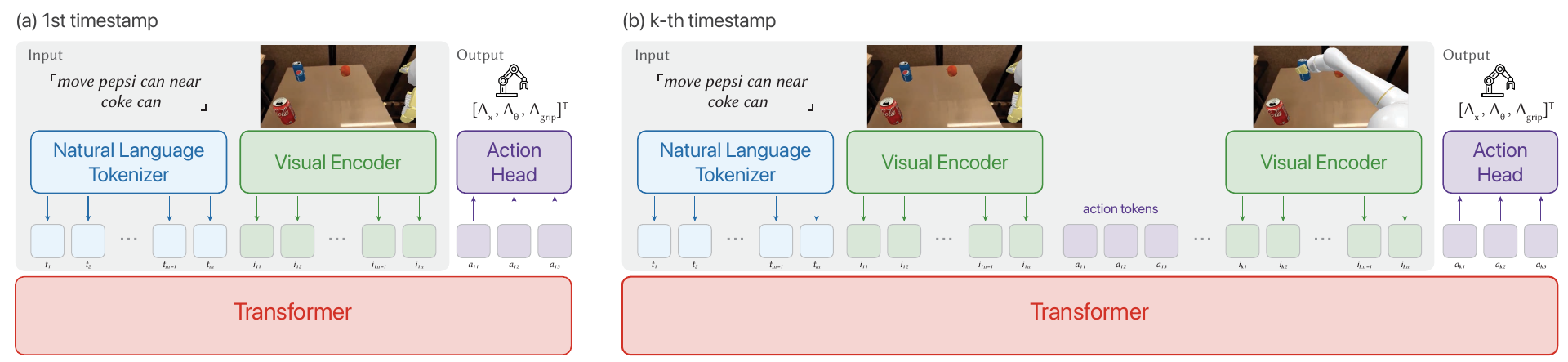}
    \vspace{-20pt}
    \caption{\textbf{Architecture and workflow of a VLA model.} (a) Generating robot actions at the first timestamp. (b) Generating robot actions at the \textit{k}-th timestamp.}
    \label{fig:vla-arch}
    \vspace{-10pt}
\end{figure}

Fig.~\ref{fig:vla-arch} illustrates the common architecture of a VLA model. Given a natural language input $T$ and an image input $I_1$ at the first timestamp, the VLA model’s natural language tokenizer and visual encoder first project $T$ and $I_1$ into sets of tokens, $\mathbf{T} = \{t_1,\dots,t_m\}$ and $\mathbf{I}_1 = \{i_{11},\dots,i_{1n}\}$, respectively. 
These tokens are then concatenated and fed into a transformer model to predict action token(s) $\mathbf{A}_1$. 
An action head layer then de-tokenizes $\mathbf{A}_1$ into robot action values $[\Delta_{x1}, \Delta_{\theta1}, \Delta_{grip1}]$, where $\Delta_{x1}\in\mathbb{R}^3$, $\Delta_{\theta1}\in\mathbb{R}^3$, and $\Delta_{grip1}\in\mathbb{R}^1$ denote the translation ($x,y,z$ movement), rotation, and gripper actions of the end-effector that should be performed after the observation, respectively.

After the robot performs these actions, the action tokens $\mathbf{A}_1$ and a new set of image tokens $\mathbf{I}_2=\{i_{21},\dots,i_{2n}\}$, based on the visual observation at the next timestamp, are concatenated with $\mathbf{T}$ and $\mathbf{I}_1$. A new input sequence $\{\mathbf{T}, \mathbf{I}_1, \mathbf{A}_1, \mathbf{I}_2\}$ is then fed into the transformer to predict new action tokens $\mathbf{A}_2$. The VLA model continues this process until the task is completed or the pre-defined maximum number of steps is exceeded.

\subsection{Training and Evaluation of the VLA Model}

\smalltitle{Training.} 
There are typically two approaches to training a VLA model: 
(1) training from scratch and 
(2) fine-tuning a general-purpose VLM. 
{\em Training from scratch} involves building a VLA model directly on robot demonstration data. 
This approach is commonly applied for models with relatively small-scale architectures and limited computational resources, such as RT-1~\cite{brohan2022rt} and Octo~\cite{team2024octo}, which typically have fewer than 100 million parameters. 
By contrast, fine-tuning from an existing VLM leverages the flexibility of re-using a larger model (more than 1 billion parameters), such as Llava~\cite{liu2024visual, kim2024openvla}, which has been pre-trained on massive amounts of image and text data from diverse domains. 
The large-scale pre-training may enhance the generalizability of the fine-tuned VLA models, particularly when manipulating unseen objects and tasks.

\smalltitle{Evaluation.} 
A VLA model is usually evaluated by measuring its performance on specific skills, such as picking up an object in a scene. 
To evaluate a particular skill, developers must first create an adequate testing scene, which involves configuring target objects, \responseline{confounding objects}, and environmental factors such as lighting conditions. Additionally, a text prompt should be meticulously crafted before performing the manipulation task. 
To assess the model’s performance, developers usually design a set of evaluation metrics.  
For instance, when performing the task ``picking up an object'', the metrics include: 
(1) whether the robot grasps the correct object, 
(2) whether the object is successfully lifted, and 
(3) whether the robot can sustain lifting the object for a short period. 
The execution of the VLA model and the robot can occur in either a simulated or real-world environment. In real-world scenarios, manual labeling is required to compute these metrics, whereas in simulated environments, the evaluation process can be automated.

\section{\method}


In this section, we first introduce the operators included in the testing scene for a robotic manipulation task. Next, we describe the algorithm used to generate a testing scene in {\method}.

\subsection{Operators} 
\begin{figure}[t]
    \centering
    \includegraphics[width=0.95\linewidth]{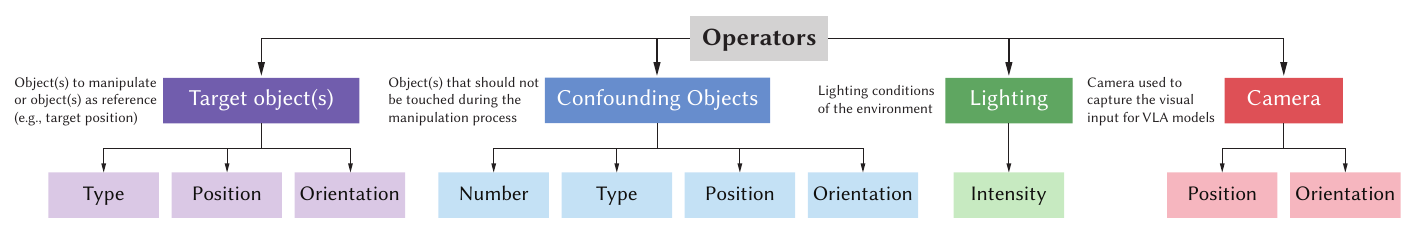}
    \vspace{-10pt}
    \caption{Operators used for testing VLA models in {\method}.}
    \label{fig:operators}
    \vspace{-15pt}
\end{figure}

We consider \textbf{four} categories of testing operators in {\method} as shown in Fig.~\ref{fig:operators}, resulting in a total of ten testing operators. \responseline{We choose these operators because they are the most essential ones in the configuration of a robotic manipulation task~\cite{siciliano2008springer, murray2017mathematical}. Thus, {\method} can be easily adapted for a wide range of tasks. Additionally, by only changing these operators, our generated scenes will not change the intrinsic nature of a manipulation task. This ensures that all test cases will not require the VLA models to perform actions that were never performed during training.}

\begin{itemize}[leftmargin=*]
    \setlength\itemsep{0.5mm}
    
    \item {\bf \em Target object(s).} A target object refers to an object that is required to be manipulated in a task (e.g., an object to pick up) or an object that serves as a reference in a task (e.g., an object to be placed on). For each target object, we consider three different operators: (1) type of object (e.g., \textit{Pepsi can}), (2) position, and (3) orientation.

    \item {\bf \em \responseline{Confounding objects}.} Different from {\em target object(s)}, \responseline{confounding objects} are typically not part of the task, and the robot is expected to avoid contacting these objects during manipulation. In addition to the three operators for {\em target object(s)}, we also consider the number of \responseline{confounding objects} as a distinct operator in the \responseline{confounding objects} category.

    \item {\bf \em Lighting.} The lighting condition affects the rendering of images captured by the camera, thereby influencing the visual input of the VLA models. We consider lighting intensity as one operator.

    \item {\bf \em Camera.} The camera’s pose can also affect the visual input of the VLA models. We consider its position and orientation as two separate operators in {\method}.
    
\end{itemize}

\subsection{Testing Scene Generation} We summarize the algorithm for generating a testing scene in Algorithm~\ref{alg:test_generation}. Intuitively, there are four distinct steps: (1) sampling semantically valid target object(s), (2) sampling \responseline{confounding objects} (optional), (3) mutating lighting conditions (optional), and (4) mutating camera poses (optional). 

\smalltitle{Sampling semantically valid target object(s).} For a robotic manipulation task, we first randomly sample target object(s) without replacement from the object database $\mathcal{O}$ (Lines 5-9). We also check if the selected objects are \textit{semantically valid} according to the task requirements (Line 4). 
{\responseref{}
For instance, in a task requiring object A to be placed over object B (Task 3), it is possible to put an apple on a plate, but it is impractical to put an apple on a ball, since placing an apple on a ball stably could be very challenging. Similarly, for Task 4, it is impossible to put an apple inside a Coke can. We consider scenes that can not be successfully manipulated by the robots to be \textit{semantically invalid}. We manually created a list of invalid objects for tasks 3 and 4 to check if a generated test scene is semantically valid.
}
After selecting the target object(s), we proceed to sample their positions and orientations (Line 7). In our empirical study, we used a random sampler for pose selection. If there are multiple target objects, we ensure they remain far enough apart ($>safe\_dist$) to avoid generating invalid testing scenes. For instance, in the previous example, if object A is already placed on object B during the sampling of their positions, the testing scene becomes invalid.

\begin{figure}[t]
\begin{algorithm}[H]
\scriptsize
\caption{Testing scene generation in {\method}.}
\label{alg:test_generation}
\setlength{\columnsep}{12pt}
\vspace{-5pt}
\begin{multicols}{2}
\KwIn{an object database $\mathcal{O}$, the robotic manipulation task $T$, the number of target objects $N_{target}$, the number of \responseline{confounding objects} $N_{confound}$, the lighting mutation option $lighting\_flag$, and the camera mutation option $camera\_flag$, the minimum distance between objects $safe\_dist$}
\KwOut{a testing scene}
$\mathcal{O}~\gets~\mathsf{random\_shuffle}(\mathcal{O}$)\;
$O_{target},O_{confound}~\gets~\emptyset,\emptyset$\;
$intensity,cam_{pos},cam_{ori}~\gets~$default values\;
\While{not $\mathsf{semantic\_valid}(O_{target}, T)$} {
    \For{$i~$in$~1,\dots,N_{target}$}{
        $obj~\gets~\mathcal{O}\mathsf{.pop()}$\;
        $pos,ori~\gets~\mathsf{pose\_sampler}(O_{target}, safe\_dist)$\;
        $O_{target}~\gets~O_{target}\bigcup(obj, pos, ori)$\;
    }
}
\If{$N_{confound}>0$}{
    \For{$i~$in$~1,\dots,N_{confound}$}{
        $obj~\gets~\mathcal{O}\mathsf{.pop()}$\;
        $pos,ori~\gets~\mathsf{pose\_sampler}(O_{target}\bigcup O_{confound},$
        $safe\_dist)$\;
        $O_{confound}~\gets~O_{target}\bigcup(obj, pos, ori)$\;
    }
}
\If{$lighting\_flag$}{
    $intensity~\gets~\mathsf{mutate\_lighting}(intensity)$\;
}
\If{$camera\_flag$}{
    $cam_{pos}, cam_{ori}~\gets~\mathsf{mutate\_camera}(cam_{pos},$
    $cam_{ori})$\;
}
\KwRet{$\{O_{target}, O_{confound}, intensity, (cam_{pos}, cam_{ori})\}$}\;
\end{multicols}
\vspace{2pt}
\end{algorithm}
\vspace{-16pt}
\end{figure}

\smalltitle{Sampling \responseline{confounding objects}.} If the number of \responseline{confounding objects} $N_{confound}$ is greater than 0 (Line 11), \responseline{we further randomly sample confounding objects from $\mathcal{O}$ without replacement} (Lines 12-16). \responseline{Note that we consider similar objects with different semantics to be different. For instance, a Coke can is considered different from a Pepsi can. Therefore, having both of them as confounding objects in one scene is possible.} For each \responseline{confounding object}, we follow the same procedure as used for sampling the target objects to assign their position and orientation (Line 14). In our empirical study, we utilized a random sampler to generate positions and orientations.

\smalltitle{Mutating lighting conditions and camera poses.} For lighting conditions, we generate a factor $\alpha$ to mutate the lighting intensity value (Lines 18-20). For the camera, we adjust its position by a distance $d$ and rotate it by an angle $\theta$ (Lines 21-23). Note that both $d$ and $\theta$ are kept small to ensure that the entire scene remains within the camera’s view. In our empirical study, we used a random generator to generate $\alpha$, $d$, and $\theta$ within specified ranges. 

\responseline{After generating each scene, we compared its configuration with existing ones to prevent duplication. Throughout our experiments, no identical test scenes were generated.}

\section{Empirical Study}

Based on {\method}, we conducted an empirical study to investigate the performance of SOTA VLA models. In this section, we first list our research questions, followed by the empirical setup, which includes subject VLA models, subject robotic manipulation tasks, prompt template, and implementation details.

\subsection{Research Questions}

Our empirical study investigates the following research questions to examine the key factors in reliable robotic manipulation: \textit{basic performance} (RQ1), \textit{task complexity} (RQ2), \textit{perception robustness} (RQ3 \& RQ 4), \textit{OOD robustness} (RQ5), \responseline{and language model robustness (RQ6)}.

\begin{itemize}[leftmargin=*]
    \setlength\itemsep{0.5mm}
    
    \item {\bf \em RQ1. \rqone} \\
    This research question seeks to evaluate the performance of SOTA VLA models in various robotic manipulation tasks. While previous studies~\cite{brohan2022rt, team2024octo} typically rely on a few hand-crafted test cases, we use {\method} to generate a large number of test cases. This approach enables a more comprehensive assessment of SOTA VLA models’ performance, providing insights into current challenges and opportunities.

    \item {\bf \em RQ2. \rqtwo} \\
    Intuitively, the presence of more \responseline{confounding objects} (\ie, objects unrelated to the assigned manipulation task) increases the complexity of a robotic manipulation task for VLA models. However, it remains unclear whether there is an upper limit to the task complexity that a VLA model can handle effectively. To address this, we conduct controlled experiments to help practitioners understand how VLA models perform as the number of \responseline{confounding objects} varies.

    \item {\bf \em RQ3. \rqthree} \\
    When deploying robotics and VLA models in real-world environments, external conditions can vary significantly. Ideally, a VLA model should be robust against different lighting conditions, such as varying illumination intensities. This research aims to offer practitioners practical guidelines on the robustness of VLA models under diverse lighting setups.

    \item {\bf \em RQ4. \rqfour} \\
    Since VLA models are pre-trained on large-scale vision datasets, they are expected to demonstrate robustness when input images are captured from various angles. This research question seeks to determine whether and to what extent current VLA models maintain robustness when operating with different camera poses.

    \item {\bf \em RQ5. \rqfive} \\
    Objects that were unseen in the robotic demonstration data may present when deploying robotics and VLA models in practical scenarios. It is unclear whether the large-scale pre-training of VLA models can make it generalizable to these unseen objects. Also, there is currently limited understanding of the performance gap between seen-object and unseen-object robotic manipulation tasks. To address this, we investigate this research question by leveraging an external object database to assess the limitations and challenges.

    {\responseref{}

    \item {\bf \em RQ6. \rqsix} \\
    The task instruction plays a vital role in the robotic manipulation with VLA models. Task instructions for a VLA model can be paraphrased using different words and sentence structures. Ideally, a VLA model is expected to perform robustly against different mutations of task instructions that carry the same meanings. In this research question, we seek to understand whether VLA models could demonstrate such {\em language model robustness}.
    }
    
\end{itemize}

\subsection{Subject VLA Models}

To investigate our research questions, we studied four different series of open-source VLA models: RT-1~\cite{brohan2022rt}, RT-1-X~\cite{padalkar2023open}, Octo~\cite{team2024octo}, and OpenVLA~\cite{kim2024openvla}. For RT-1~\cite{brohan2022rt}, we studied three publicly available variants: \textit{RT-1-1k}, \textit{RT-1-58k}, and \textit{RT-1-400k}~\footnote{The number denotes the training steps taken for training an RT-1 model.}. We also studied two variants for Octo~\cite{team2024octo}: \textit{Octo-small} and \textit{Octo-base}. Note that we were unable to study the SOTA VLA model, RT-2~\cite{brohan2023rt}, as it has not yet been made publicly available. Thus, we included a total of seven subject VLA models in our empirical study. Table~\ref{tab:subject_vlas} summarizes the characteristics of these models, and we provide more details for each model below.

\begin{table}[t]
    \renewcommand{\arraystretch}{1}
    \caption{Overview of subject VLA models.}
    \label{tab:subject_vlas}
    \vspace{-10pt}
    \scriptsize
    \centering
    \begin{tabular}{|l|c|c|c|c|c|c|c|}
        \hline
        \multirow{2}{*}{\textbf{VLA Models}} & \multicolumn{3}{c|}{\textbf{RT-1}} & \multirow{2}{*}{\textbf{RT-1-X}} & \multirow{2}{*}{\textbf{Octo-small}} & \multirow{2}{*}{\textbf{Octo-base}} & \multirow{2}{*}{\textbf{OpenVLA-7b}} \\
        \cline{2-4}
         & {1k} & {58k} & {400k} & & & & \\
        \hline
        \hline
        \textbf{Model Size} & \multicolumn{3}{c|}{35M} & 35M & 27M & 93M & 7.6B\\
        \hline
        \textbf{Release Date} & \multicolumn{3}{c|}{Dec. 2022} & Oct. 2023 & Dec. 2023 & Dec. 2023 & Jun. 2024\\
        \hline
        \textbf{\responseline{Report Performance}} & \responseline{---} & \responseline{---} & \responseline{92\%} & \responseline{73\%} & \responseline{30\%} & \responseline{71\%} & \responseline{71\%} \\
        \cline{2-8}
        \responseline{(Dataset)} & \multicolumn{4}{c|}{\responseline{RT-1 Paper 6 Skills}} & \multicolumn{2}{c|}{\responseline{UR5}} & \responseline{WidowX}\\
        \hline
    \end{tabular}
    \vspace{-15pt}
\end{table}

\begin{itemize}[leftmargin=*]
    \setlength\itemsep{0.5mm}
    
    \item \textbf{RT-1}~\cite{brohan2022rt} model was released by Google Research. The RT-1 model includes a FiLM EfficientNet-B3~\cite{perez2018film, tan2019efficientnet} pre-trained on the ImageNet dataset~\cite{russakovsky2015imagenet} for tokenizing visual inputs and language instructions before connecting to a transformer to generate robotic manipulation actions. The RT-1 model was trained on an unpublished set of 130k robot demonstrations collected by Google.

    \item \textbf{RT-1-X}~\cite{padalkar2023open} model was released by Google DeepMind. RT-1-X shares the same model architecture as RT-1. However, RT-1-X was trained on the open-source dataset, Open X-Embodiment~\cite{padalkar2023open}, which includes 160k robot demonstrations collected from 22 different robots.

    \item \textbf{Octo}~\cite{team2024octo} model, released by UC Berkeley, includes a ViT model~\cite{dosovitskiy2020image} as its backbone transformer. It introduces the ``readout’’ token, allowing developers to flexibly add new observation inputs or action output heads to the model during downstream fine-tuning. Octo was trained on a subset of the Open X-Embodiment dataset, which includes about 65k robot demonstrations. Octo-small (27M parameters) and Octo-base (93M parameters) are two variants of Octo, utilizing ViT-S and ViT-B as their backbone transformers, respectively.

    \item \textbf{OpenVLA-7b}~\cite{kim2024openvla} is the most recent VLA model released by Stanford University. OpenVLA-7b includes a 600M-parameter visual encoder consisting of pre-trained SigLIP~\cite{zhai2023sigmoid} and DinoV2~\cite{oquabdinov2} models and a 7B-parameter Llama 2~\cite{touvron2023llama} as the language model backbone. OpenVLA-7b was then fine-tuned on the Open X-Embodiment dataset.

\end{itemize}

\subsection{Robotic Manipulation Tasks}

We included four different robotic manipulation tasks in our study. Now, we briefly introduce each of them. We also refer to the supplementary materials for the video demo of each task.

\begin{itemize}[leftmargin=*]
    \setlength\itemsep{0.5mm}
    
    \item {\bf \em Task 1: Pick up an object.} This task requires a VLA model to identify the target object and generate the appropriate control signals to grasp and lift it. To succeed, the robot must grasp the correct object and lift it at least 0.02 meters for five consecutive frames.

    \item {\bf \em Task 2: Move object A to object B.} This task requires a VLA model to first identify the source object (A) and then output the corresponding control signals to move it near the target object (B). To succeed, the robot must move the correct object to within 0.05 meters of the target object.

    \item {\bf \em Task 3: Put object A on top of object B.} Different from Task 2, this task requires the VLA model to output control signals that could stack object A on top of object B. To succeed, object A should be placed stably on top of object B.

    \item {\bf \em Task 4: Put object A into object B.} Different from Task 3, this task requires the VLA model to generate control signals that place object A inside object B (e.g., into a kitchen sink or a basket). To succeed, object A must be completely inside object B.
    
\end{itemize}

\responseline{For each task, {\method} automatically verifies whether a VLA model has successfully performed it by checking if all corresponding specifications (e.g., the lifted object’s height) are met.}

\subsection{Prompt Templates} 
\label{subsec:prompt}

For each task in RQ1~\textasciitilde~RQ5, we follow the previous work~\cite{brohan2022rt, padalkar2023open, kim2024openvla, li2024evaluating} to use the standard prompt template for each task: (1) {\small \texttt{pick up [object name]}}, (2) {\small \texttt{move [object name] near [object name]}}, (3) {\small \texttt{put [object name] on [object name]}}, and (4) {\small \texttt{put [object name] into [object name]}}. \responseline{In RQ6, we use the prompts mutated from these standard templates.}


\subsection{Implementation Details} 

We conducted all experiments on a server with an AMD 5955WX CPU and two NVIDIA RTX A6000 GPUs. The operating system is 64-bit Ubuntu 20.04 LTS with Python 3.10 and CUDA 12.2. We implemented {\method} with Maniskill2 simulation environments~\cite{gu2023maniskill2, li2024evaluating}. We used two object databases: (1) the default object database in Maniskill2 ($N=18$) in RQ1~\textasciitilde~RQ4 and RQ6, and (2) YCB object database ($N=56$) in RQ5. The average execution time of one manipulation task was around 19.8 seconds. Our empirical study took about \responseline{586 GPU hours} in total.

\section{Results}

\subsection{RQ1: \RQone}
\label{subsec:rq1}

To investigate this research question, we leveraged {\method} to generate 1,000 scenes for each of the four robotic manipulation tasks by randomly selecting target object(s) and sampling 0 to 3 \responseline{confounding objects}. We also randomly assigned poses to these objects. To avoid collision overlaps, we maintained a minimum distance of 0.15 meters between objects during the assignment. The default lighting setups and camera poses were used for this RQ. 

\begin{table}[t]
    \renewcommand{\arraystretch}{1}
    \centering
    \scriptsize
    \caption{\setlength\fboxsep{1pt} (\textbf{RQ1}) Performance of seven subject VLA models on different manipulation tasks. The \colorbox{red1}{\textcolor{white}{top-1}}, \colorbox{red4}{top-2}, and \colorbox{lightred}{top-3} success rates for each step and task are highlighted, respectively.}
    \vspace{-10pt}
    \label{tab:performance}
    \begin{tabular}{|l||rrr|rrr|rrr|rrr|}
         \hline
         \multirow{2}{*}{\textbf{VLA Models}} & \multicolumn{3}{c|}{\textbf{Task 1: Pick Up}} & \multicolumn{3}{c|}{\textbf{Task 2: Move Near}} & \multicolumn{3}{c|}{\textbf{Task 3: Put On}} & \multicolumn{3}{c|}{\textbf{Task 4: Put In}} \\
         \cline{2-13}
         & \multicolumn{1}{c}{Grasp} & \multicolumn{1}{c}{Lift} & \multicolumn{1}{c|}{Success} & \multicolumn{1}{c}{Grasp} & \multicolumn{1}{c}{Move} & \multicolumn{1}{c|}{Success} & \multicolumn{1}{c}{Grasp} & \multicolumn{1}{c}{Move} & \multicolumn{1}{c|}{Success} & \multicolumn{1}{c}{Grasp} & \multicolumn{1}{c}{Move} & \multicolumn{1}{c|}{Success} \\
         \hline
         \hline
         RT-1-1k & 10.1\% & 1.2\% & 0.7\% & 9.6\% & 3.2\% & 1.4\% & 0.4\% & 0.0\% & 0.0\% & 1.1\% & 0.0\% & 0.0\%  \\
         RT-1-58k & \cellcolor{red4}44.5\% & \cellcolor{red4}32.6\% & \cellcolor{red4}28.2\% & \cellcolor{red4}38.6\% & \cellcolor{red4}25.2\% & \cellcolor{red4}10.9\% & 0.3\% & 0.0\% & 0.0\% & 1.3\% & 0.0\% & 0.0\% \\
         RT-1-400k & \cellcolor{red1}\textcolor{white}{48.4\%} & \cellcolor{red1}\textcolor{white}{41.0\%} & \cellcolor{red1}\textcolor{white}{34.4\%} & \cellcolor{lightred}38.8\% & \cellcolor{lightred}23.7\% & \cellcolor{lightred}9.4\% & 10.4\% & 1.3\% & 0.5\% & 8.9\% & 0.1\% & 0.1\% \\
         RT-1-X & \cellcolor{lightred}34.0\% & \cellcolor{lightred}25.8\% & \cellcolor{lightred}19.5\% & 25.4\% & 15.2\% & 5.8\% & \cellcolor{red1}\textcolor{white}{17.2\%} & \cellcolor{red1}\textcolor{white}{4.4\%} & \cellcolor{red1}\textcolor{white}{2.3\%} & 17.8\% & 0.7\% & 0.4\% \\
         Octo-small & 9.0\% & 2.0\% & 0.8\% & 14.8\% & 4.1\% & 1.5\% & \cellcolor{red4}27.5\% & \cellcolor{red4}4.6\% & \cellcolor{red4}2.2\% & \cellcolor{red1}\textcolor{white}{34.6\%} & \cellcolor{red1}\textcolor{white}{1.1\%} & \cellcolor{red1}\textcolor{white}{1.1\%} \\
         Octo-base & 2.3\% & 0.4\% & 0.0\% & 4.9\% & 1.5\% & 0.6\% & 19.1\% & 2.6\% & 1.2\% & \cellcolor{red4}32.5\% & \cellcolor{red4}1.1\% & \cellcolor{red4}1.1\%  \\
         OpenVLA-7b & 15.1\% & 7.1\% & 5.9\% & \cellcolor{red1}\textcolor{white}{43.0\%} & \cellcolor{red1}\textcolor{white}{23.3\%} & \cellcolor{red1}\textcolor{white}{12.7\%} & \cellcolor{lightred}36.8\% & \cellcolor{lightred}5.4\% & \cellcolor{lightred}2.1\% & \cellcolor{lightred}19.9\% & \cellcolor{lightred}1.1\% & \cellcolor{lightred}1.1\% \\
         \hline
         \hline
         Avg. & 23.3\% & 15.7\% & 12.8\% & 25.0\% & 13.7\% & 6.0\% & 16.0\% & 2.6\% & 1.2\% & 16.6\% & 0.6\% & 0.5\% \\
        \hline
    \end{tabular}
    \vspace{-15pt}
\end{table}

Table~\ref{tab:performance} presents the performance of seven VLA models on these tasks and scenes. Overall, we find that current VLA models do not perform well in the four studied robotic manipulation tasks. The average success rates across the seven VLA models for the four tasks are 12.4\%, 6.0\%, 1.2\%, and 0.5\%, respectively. In Task 1, the best-performing VLA model, RT-1-400k, succeeded in only 34.4\% of the test scenes. We also observed significant performance drops in Task 2, Task 3, and Task 4 compared to Task 1. The best-performing VLA model in Task 2, OpenVLA-7b, completed only 12.7\% of the test scenes. In Task 3 and Task 4, the best-performing VLA models (RT-1-X and Octo-small) achieved success rates of just 2.2\% and 2.1\%, respectively. One possible reason is that these three tasks are more complex than Task 1, as they require multiple steps of reasoning (i.e., identifying the source and target objects and their positions before generating the corresponding control signals). Among the seven VLA models studied, we did not find any model that performed significantly better than the others across different tasks.

\begin{finding}
\label{findings:overall_performance}
    All VLA models exhibit \textbf{subpar performance} in the four studied robotic manipulation tasks, particularly in those that require identifying multiple target objects (i.e., Task 2, Task 3, and Task 4). These results suggest that the development of VLA models is still in its early stages, as they are far from being ready for deployment in real-world scenarios.
\end{finding}
\vspace{-1ex}

To understand the rationale behind the subpar performance of these VLA models, we broke down each testing scene according to the different steps required to complete a task successfully. Specifically, we measured the success rates of each step in a testing scene for the four tasks, as presented in Table~\ref{tab:performance}. For instance, Task 1 requires the robot to (1) grasp the target object (the column \textit{Grasp} in Table~\ref{tab:performance}), (2) lift the object significantly (the column \textit{Lift} in Table~\ref{tab:performance}), and (3) continue lifting for five consecutive frames (the column \textit{Success} in Table~\ref{tab:performance}). In all four tasks, we found that the success rates dropped significantly between steps. For example, in Task 1, the average success rate of correctly grasping the target object is 23.3\%. The success rates for lifting and maintaining the lift then drop to 15.7\% and 12.4\%, respectively. These results indicate that the current VLA models struggle to interpret natural language instructions that require the robot to perform multiple sequential actions.

In Task 2, Task 3, and Task 4, VLA models successfully picked up source objects in 16\% to 25\% of scenes. However, they then failed to identify the target objects, with only 0.6\% to 13.7\% of objects being correctly moved and 0.5\% to 6.0\% of objects being correctly placed. To improve, one could consider the idea of {\em chain-of-thought} prompting in LLMs. Specifically, the complex task instructions could be decoupled into individual action steps, allowing the VLA model to be prompted step-by-step.

\begin{finding}
\label{findings:complex_command}
    Current VLA models cannot successfully execute tasks that require multiple steps of action, highlighting the urgent need to improve their capabilities in accurately interpreting task instructions.
\end{finding}
\vspace{-1ex}

{\responseref{}

\textbf{\textit{Testing Sufficiency.}} To evaluate the quality of the test scenes generated by {\method}, we further calculated test coverage. However, to the best of our knowledge, no existing coverage metrics are applicable to either VLA models or robotic manipulation. Metrics such as neuron coverage could be computationally intensive to calculate, as VLA models are typically trained on very large datasets with millions or even billions of parameters.

\begin{wraptable}{r}{0.4\linewidth}
\vspace{-12pt}
\renewcommand{\arraystretch}{1}
\vspace{-3pt}
\caption{\responseref{} Trajectory coverage results with different number of test cases ($n$).}
\vspace{-10pt}
\label{tab:coverage}
\scriptsize
\centering
\begin{tabular}{|l|c|c|c|c|}
\hline
$n$ & \textbf{Task 1} & \textbf{Task 2} & \textbf{Task 3} & \textbf{Task 4} \\
\hline
\hline
10 & 0.09 & 0.09 & 0.10 & 0.10 \\
100 & 0.64 & 0.66 & 0.63 & 0.62 \\
1000 & 1.00 & 1.00 & 1.00 & 1.00 \\
\hline
\end{tabular}
\vspace{-15pt}
\end{wraptable}We followed previous work in autonomous driving to implement trajectory coverage~\cite{hu2021coverage}. We believe this is a pragmatic choice for enabling test coverage measurements. We acknowledge its limitations in Sec.~\ref{sec:threats}. Specifically, we calculated how many novel target object positions were covered {\em w.r.t.} the size of the manipulation platform (i.e., the desk). As shown in Table~\ref{tab:coverage}, we found that increasing the number of generated test cases $n$ from 10 to 1000 allowed {\method} to achieve 100\% coverage across all four tasks, indicating that our generated test scenes are sufficient.

}

\subsection{RQ2: \RQtwo} 

\begin{figure}[t]
    \centering
    \includegraphics[width=0.95\linewidth]{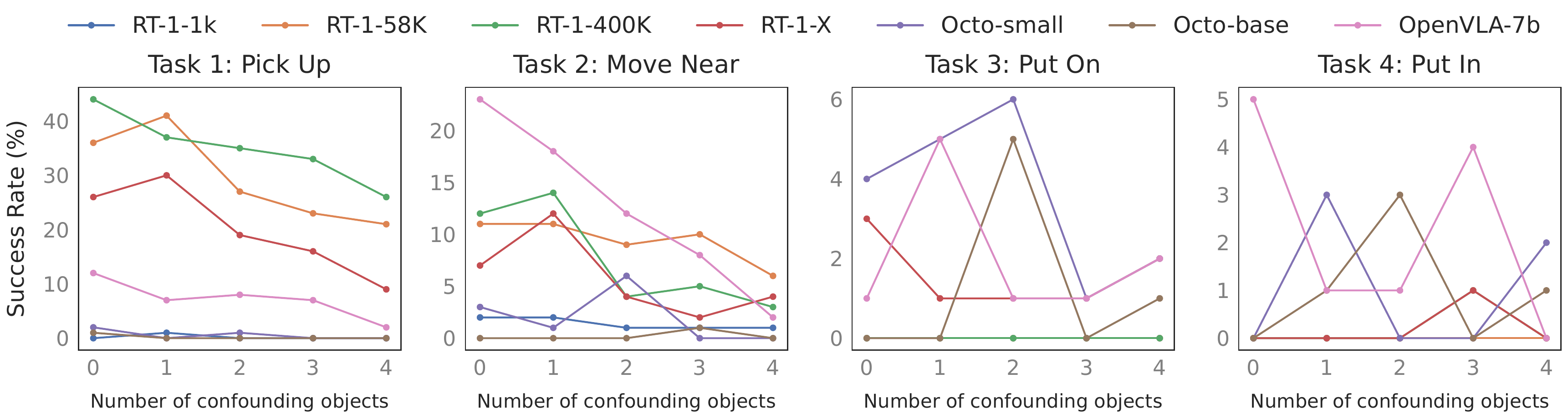}
    \vspace{-10pt}
    \caption{(\textbf{RQ2}) VLA performance vs. the number of \responseline{confounding objects}.}
    \label{fig:num_obstacles}
    \vspace{-19pt}
\end{figure}

To investigate RQ2, we used {\method} to generate 100 scenes for each task with a fixed number of $n$ \responseline{confounding objects}, where $n$ was set to 0, 1, 2, 3, and 4, resulting in a total of 500 scenes per task. We used the default lighting conditions and camera poses when investigating this RQ.

The results are depicted in Fig.~\ref{fig:num_obstacles}. In the first two tasks (i.e., Task 1 and Task 2), we observed that the VLA model’s success rate decreased as the number of \responseline{confounding objects} increased. The average success rates across different VLA models and scenes dropped from 17.3\% to 8.3\% and from 8.3\% to 1.1\% when increasing the number of \responseline{confounding objects} from 0 to 4 in Task 1 and Task 2, respectively. However, in Task 3 and Task 4, we did not find a similar pattern. This may be because all VLA models performed poorly even without any \responseline{confounding objects} ($n=0$). The average success rates across different VLA models and scenes were 1.2\% and 0.7\% when $n=0$ and 1.1\% and 0.4\% when $n=4$ for Task 3 and Task 4, respectively.

\begin{finding}
\label{findings:obstacles}
    The number of \responseline{confounding objects} affects the VLA model’s performance, indicating that VLA models become unreliable in more complex environments. {When there were four \responseline{confounding objects} presented in the scene, the VLA models only passed 8.2\%, 2.3\%, 1.0\%, and 0.4\% of the test scenes in Task 1, Task 2, Task 3, and Task 4, respectively.}
\end{finding}
\vspace{-1.5ex}

\begin{figure}[t]
    \centering
    \includegraphics[width=0.95\linewidth]{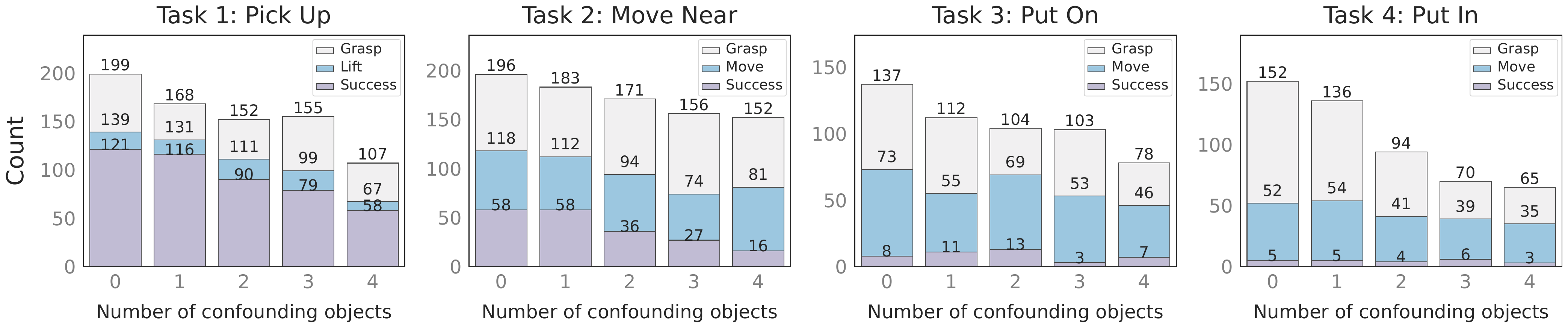}
    \vspace{-12pt}
    \caption{(\textbf{RQ2}) The number of successful scenes at different steps vs. the number of \responseline{confounding objects}. 
    }
    \label{fig:step_num_obstacles}
    \vspace{-12pt}
\end{figure}

Similar to RQ1, we also analyzed each testing scene and its success at each individual step. We counted the number of testing scenes that succeeded at each step and presented the results in Fig.~\ref{fig:step_num_obstacles}. We found that the major challenge faced by the current VLA models lies in their inability to identify the correct object to manipulate.

In all four tasks, we found that the success rates for grasping the target object decreased as the number of \responseline{confounding objects} increased. These results indicate that when there are multiple \responseline{confounding objects}, the VLA models struggle to locate the correct object to manipulate. Specifically, without any \responseline{confounding objects}, the VLA models successfully located the object to manipulate in 199, 196, 137, and 152 out of 700 cases for Task 1, Task 2, Task 3, and Task 4, respectively. However, with four \responseline{confounding objects}, the VLA models successfully located the correct object in only 107, 152, 78, and 65 out of 700 cases for the four tasks, respectively. The average success rates of grasping across VLA models dropped to 53.7\%, 77.6\%, 47.4\%, and 42.8\% for four tasks, respectively.

\begin{finding}
\label{findings:obstacles_identification}
    With multiple \responseline{confounding objects} in the scene, the VLA models’ ability to accurately locate the correct object to manipulate can be significantly affected.
\end{finding}
\vspace{-1.5ex}

\begin{wrapfigure}{r}{0.36\textwidth}
     \centering
     \vspace{-12pt}
     \begin{subfigure}[b]{0.36\textwidth}
         \centering
         \includegraphics[width=1\linewidth]{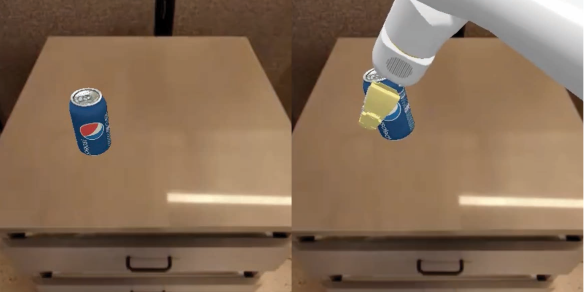}
         \vspace{-15pt}
        \caption{No \responseline{confounding objects}}
        \vspace{-5pt}
         \label{fig:grasp_succ}
     \end{subfigure}%
     \vspace{3mm}
     \begin{subfigure}[b]{0.36\textwidth}
         \centering
         \includegraphics[width=1\linewidth]{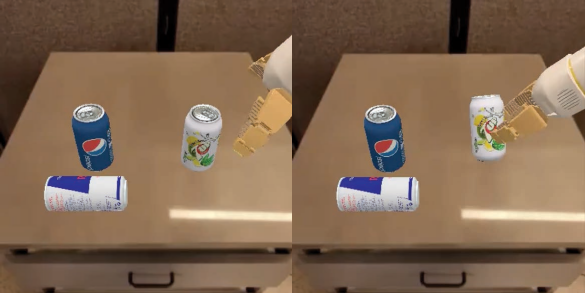}
         \vspace{-15pt}
        \caption{Two \responseline{confounding objects}}
        \vspace{-5pt}
         \label{fig:grasp_fail}
     \end{subfigure}%
     \vspace{-3pt}
    \caption{RT-1-X perform Task 1 with different number of \responseline{confounding objects}.}
     \vspace{-12pt}
    \label{fig:grasp_case}
\end{wrapfigure}Fig.~\ref{fig:grasp_case} shows two different testing scenes of Task 1. In both scenes, the VLA model, RT-1-X, was asked to pick up the {\em Pepsi can}. When there was no \responseline{confounding object} (Fig.~\ref{fig:grasp_succ}), RT-1-X successfully grasped and lifted the Pepsi can. However, when there were two \responseline{confounding objects} (i.e., {\em 7 Up can} and {\em Red Bull can}), RT-1-X failed to locate the Pepsi can and eventually picked up the 7 Up Can.


\vspace{1mm}
\textbf{\textit{The Impact of Confounding Objects' Similarity.}} In addition to the number of confounding objects, another factor that may impact the performance of VLA models is the similarity between the confounding object(s) and the target object(s). We manually examined our object database and identified two groups of similar objects: (1) pop cans of different brands and (2) cubes of different colors. Then, we categorized our experiment results from RQ2 into two groups: (1) cases where the target object(s) and confounding object(s) were similar and (2) cases where they were dissimilar. We compared the success rates of these two groups under different $n$ across the four tasks. Specifically, we conducted Mann-Whitney U tests, which showed no significant performance difference between the two groups, with $p$-values of $0.443$, $0.614$, $0.657$, and $0.443$ for the four tasks, respectively. Effect sizes are $0.291$, $0.234$, $0.271$, and $0.257$. These results indicate that the similarity between the target objects and the confounding object(s) has little impact on the VLA model's performance.


\subsection{RQ3: \RQthree}

To answer this research question, we first collected all the successfully executed scenes for each VLA model in RQ1, resulting in a total of 1,434 passed test cases. We then re-executed these test cases three times after randomly increasing or decreasing the lighting intensities ($N=4,302$). 
Specifically, we randomly sampled a factor $\alpha \in (1, 20]$ for increasing or $\alpha \in [1/20, 1)$ for decreasing the lighting intensity. This factor was then multiplied by the default lighting intensity used in RQ1. We manually checked the cases when $\alpha$ was set to $20$ and $1/20$ and confirmed that the images captured under such lighting conditions were still recognizable by humans.

We present the experiment results in Table~\ref{tab:lighting}. Overall, we found that randomly perturbing the lighting conditions significantly affected the performance of the VLA models. Out of 1,434 passed test cases with default lighting conditions, only 878.4 test cases (average results over three mutations for each test case) could still be successfully executed with perturbed lighting conditions. The perturbed lighting conditions had a significant impact on Task 1, Task 3, and Task 4, where only about half of the test cases could still be passed after the perturbation. In Task 2, however, five out of seven VLA models exhibited robust performance despite changes in lighting conditions.

\begin{table}[t]
    \renewcommand{\arraystretch}{1}
    \centering
    \scriptsize
    \caption{\setlength\fboxsep{1pt} (\textbf{RQ3}) Performance of subject VLA models under default (\textbf{Def.}) and mutated (\textbf{Mut.}) lighting conditions. Each cell represents the number of successfully passed test scenes by different VLA models in different tasks. The \colorbox{red1}{\textcolor{white}{top-1}}, \colorbox{red4}{top-2} and \colorbox{lightred}{top-3} robust VLA models are highlighted, respectively.
    }
    \vspace{-10pt}
    \label{tab:lighting}
    \begin{threeparttable}
    \begin{tabular}{|l||rr|rr|rr|rr||rrc|}
         \hline
         \multirow{3}{*}{\textbf{VLA Models}} & \multicolumn{2}{c|}{\textbf{Task 1}} & \multicolumn{2}{c|}{\textbf{Task 2}} & \multicolumn{2}{c|}{\textbf{Task 3}} & \multicolumn{2}{c||}{\textbf{Task 4}} & \multicolumn{3}{c|}{\multirow{2}{*}{\textbf{Overall}}} \\
         & \multicolumn{2}{c|}{\textbf{Pick Up}} & \multicolumn{2}{c|}{\textbf{Move Near}} & \multicolumn{2}{c|}{\textbf{Put On}} & \multicolumn{2}{c||}{\textbf{Put In}} & \multicolumn{3}{c|}{}\\
         \cmidrule{2-12}
         & \multicolumn{1}{c}{Def.} & \multicolumn{1}{c|}{Mut.$^\dagger$} & \multicolumn{1}{c}{Def.} & \multicolumn{1}{c|}{Mut.$^\dagger$} & \multicolumn{1}{c}{Def.} & \multicolumn{1}{c|}{Mut.$^\dagger$} & \multicolumn{1}{c}{Def.} & \multicolumn{1}{c||}{Mut.$^\dagger$} & \multicolumn{1}{c}{Def.} & \multicolumn{1}{c}{Mut.$^\dagger$} & \multicolumn{1}{c|}{Mut./Def. (\%)} \\
         \hline
         \hline
         RT-1-1k & 7 & 0.7 & 14 & 14.0 & 0 & --- & 0 & --- & \cellcolor{red4}21 & \cellcolor{red4}14.7 & \cellcolor{red4}70.0\% \\
         RT-1-58k & 282 & 142.0 & 109 & 109.0 & 0 & --- & 0 & --- & 491 & 251.0 & 51.1\% \\
         RT-1-400k & 344 & 163.0 & 94 & 92.0 & 5 & 5.0 & 1 & 1.0 & \cellcolor{lightred}444 & \cellcolor{lightred}261.0 & \cellcolor{lightred}58.8\% \\
         RT-1-X & 195 & 77.3 & 58 & 57.0 & 22 & 20.7 & 4 & 3.0 & 279 & 158.0 & 56.6\% \\
         Octo-small & 8 & 0.0 & 15 & 2.3 & 21 & 6.0 & 10 & 1.0 & 54 & 9.3 & 17.2\% \\
         Octo-base & 0 & --- & 6 & 1.7 & 11 & 3.0 & 11 & 0.7 & 28 & 5.4 & 19.3\% \\
         OpenVLA-7b & 59 & 21.0 & 127 & 127.0 & 20 & 20.0 & 11 & 11.0 & \cellcolor{red1}\textcolor{white}{217} & \cellcolor{red1}\textcolor{white}{169.0} & \cellcolor{red1}\textcolor{white}{77.9\%} \\
         \hline
         \hline
         Tot. & 895 & 404.0 & 423 & 403.0 & 79 & 54.7 & 37 & 16.7 & 1434 & 878.4 & 61.3\%\\
         \hline
    \end{tabular}
    \begin{tablenotes}
        \item $\dagger$ Averaged results over three mutations.
    \end{tablenotes}
    \end{threeparttable}
    \vspace{-14pt}
\end{table}

Among the seven VLA models, we found that OpenVLA-7b was the most robust, with 77.9\% of the test cases still being passed under mutated lighting conditions. A plausible explanation is that OpenVLA-7b leverages two vision models (i.e., SigLIP~\cite{zhai2023sigmoid} and DinoV2~\cite{oquabdinov2}) pre-trained on a diverse set of images and fine-tuned on the LLaVA 1.5 data mixture (1 million images and texts). As a result, OpenVLA-7b can effectively interpret images captured under varying lighting conditions. In contrast, Octo-small and Octo-base were trained solely on images captured from robot demonstrations, making them less robust when faced with lighting conditions different from the default ones. Only 17.2\% and 19.3\% of the test cases could still be successfully executed after changing the lighting conditions, respectively.

\begin{finding}
    \label{finding:lighting_model}
    Overall, the VLA models are not sufficiently robust against lighting changes. Only 61.3\% of the test cases could still be successfully executed under mutated lighting conditions. Among the seven VLA models, OpenVLA-7b demonstrated the highest robustness to lighting changes. 
\end{finding}
\vspace{-1.5ex}

\begin{figure}[h]
     \centering
     \vspace{-10pt}
     \begin{subfigure}[b]{0.4\textwidth}
         \centering
         \includegraphics[width=1\linewidth]{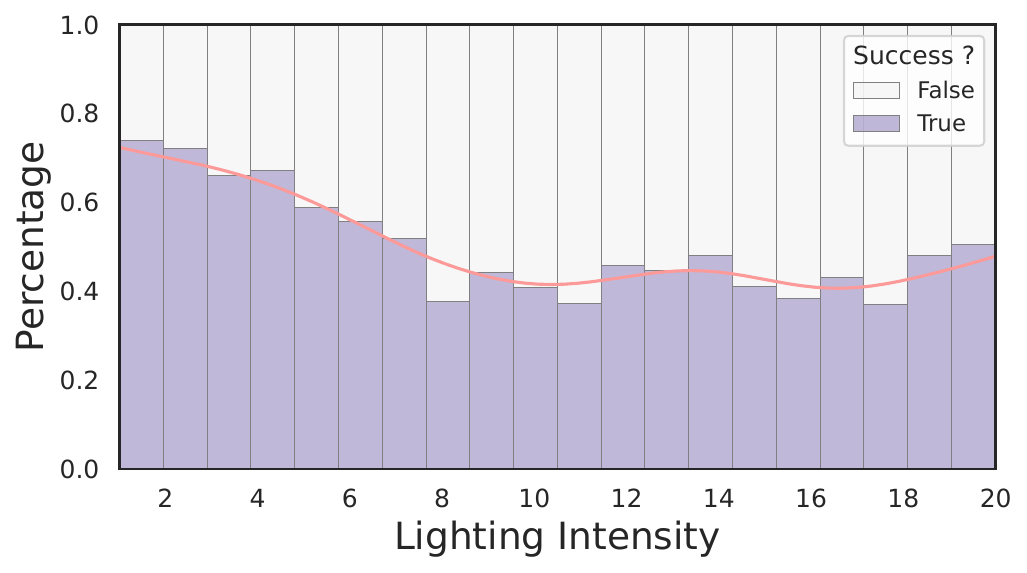}
         \vspace{-18pt}
        \caption{Increasing the lighting intensity}
        \vspace{-5pt}
         \label{fig:increase_intensity}
     \end{subfigure}%
     \hspace{10mm}
     \begin{subfigure}[b]{0.4\textwidth}
         \centering
         \includegraphics[width=1\linewidth]{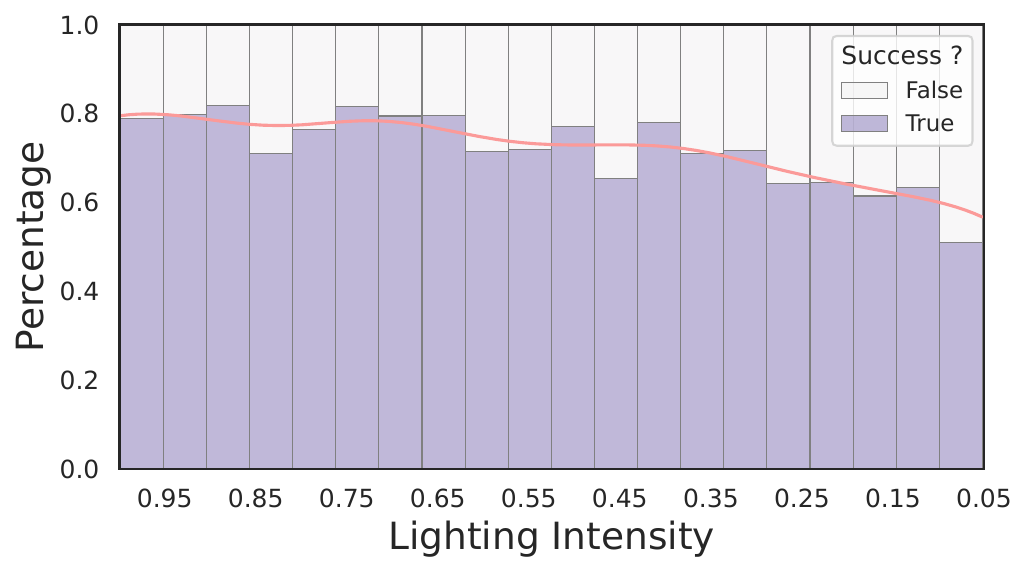}
         \vspace{-18pt}
        \caption{Decreasing the lighting intensity}
        \vspace{-5pt}
         \label{fig:decrease_intensity}
     \end{subfigure}%
     \vspace{-5pt}
    \caption{VLA performance vs. different lighting intensities.}
     \vspace{-15pt}
    \label{fig:lighting_intensity}
\end{figure}

We further investigated the extent to which the VLA models’ performance could be affected by increasing or decreasing lighting intensities. Fig.~\ref{fig:increase_intensity} and Fig.~\ref{fig:decrease_intensity} present the results.

\smalltitle{Increasing the lighting intensity ($\alpha > 1$).} Even when the lighting intensity is increased by less than 2.5 times (i.e., $\alpha < 2.5$) of the default value, the success rate of the VLA models immediately drops to around 0.7. These results indicate that even a small increase in lighting intensity can significantly degrade the performance of the VLA models. As the factor $\alpha$ increases, the success rate of the VLA models continues to decrease. In extreme cases (e.g., $\alpha > 7.5$), the VLA models succeed in less than half of the scenes.

\begin{finding}
    \label{finding:increase_lighting}
    The performance of the VLA models degrades with increasing lighting intensity. When the lighting intensity exceeds 8 times the default value, the VLA models succeed in only about 40\% of the testing scenes that could be passed under default lighting conditions.
\end{finding}
\vspace{-2ex}

\smalltitle{Decreasing the lighting intensity ($\alpha < 1$).} Similar to increasing the lighting intensity, the VLA models’ performance is also affected by even a small perturbation (e.g., $\alpha > 0.9$) when decreasing the intensity. The VLA models’ performance degrades with decreasing $\alpha$. However, compared to increasing the lighting intensity, decreasing it has less significant effects on the models’ performance. When $\alpha < 0.2$, the VLA models can still pass around 60\% of the testing scenes.

\begin{finding}
    \label{finding:decrease_lighting}
    The effect of decreasing the lighting intensity is less significant than increasing it on the VLA models’ performance. {Even when the lighting intensity is reduced to 0.15 of its default value, the VLA models can still pass 60\% of the test cases.} 
\end{finding}
\vspace{-2ex}

\subsection{RQ4: \RQfour}
Similar to RQ3, we re-executed the 1,434 passed test cases collected from RQ1 three times after randomly moving and rotating the camera from its default position ($N=4,302$). Specifically, we rotated the camera around each axis by an angle randomly sampled between $-5^{\circ}$ and $5^{\circ}$. Additionally, we randomly moved the camera away from its center by a distance randomly sampled between 0 and 5 cm. We manually checked the corner cases where the camera was rotated by $5^\circ$ and moved by 5 cm, and confirmed that the images captured with these camera angles still included the entire scene and objects. We report our experiment results in Table~\ref{tab:camera}

\begin{table}[t]
    \renewcommand{\arraystretch}{1}
    \centering
    \scriptsize
    \caption{\setlength\fboxsep{1pt} (\textbf{RQ4}) Performance of subject VLA models with default (\textbf{Def.}) and mutated (\textbf{Mut.}) camera poses. Each cell represents the number of successfully passed test scenes by different VLA models in different tasks. The \colorbox{red1}{\textcolor{white}{top-1}}, \colorbox{red4}{top-2} and \colorbox{lightred}{top-3} robust VLA models are highlighted, respectively.}
    \vspace{-10pt}
    \label{tab:camera}
    \begin{threeparttable}
    \begin{tabular}{|l||rr|rr|rr|rr||rrc|}
         \hline
         \multirow{3}{*}{\textbf{VLA Models}} & \multicolumn{2}{c|}{\textbf{Task 1}} & \multicolumn{2}{c|}{\textbf{Task 2}} & \multicolumn{2}{c|}{\textbf{Task 3}} & \multicolumn{2}{c||}{\textbf{Task 4}} & \multicolumn{3}{c|}{\multirow{2}{*}{\textbf{Overall}}} \\
         & \multicolumn{2}{c|}{\textbf{Pick Up}} & \multicolumn{2}{c|}{\textbf{Move Near}} & \multicolumn{2}{c|}{\textbf{Put On}} & \multicolumn{2}{c||}{\textbf{Put In}} & \multicolumn{3}{c|}{}\\
         \cline{2-12}
         & \multicolumn{1}{c}{Def.} & \multicolumn{1}{c|}{Mut.$^\dagger$} & \multicolumn{1}{c}{Def.} & \multicolumn{1}{c|}{Mut.$^\dagger$} & \multicolumn{1}{c}{Def.} & \multicolumn{1}{c|}{Mut.$^\dagger$} & \multicolumn{1}{c}{Def.} & \multicolumn{1}{c||}{Mut.$^\dagger$} & \multicolumn{1}{c}{Def.} & \multicolumn{1}{c}{Mut.$^\dagger$} & \multicolumn{1}{c|}{Mut./Def. (\%)} \\
         \hline
         \hline
         RT-1-1k & 7 & 0.7 & 14 & 4.3 & 0 & --- & 0 & --- & 21 & 5.0 & 23.8\% \\
         RT-1-58k & 282 & 121.3 & 109 & 27.0 & 0 & --- & 0 & --- & \cellcolor{lightred}491 & \cellcolor{lightred}39.3 & \cellcolor{lightred}30.2\% \\
         RT-1-400k & 344 & 180.3 & 94 & 21.3 & 5 & 0.0 & 1 & 0.7 & \cellcolor{red1}\textcolor{white}{444} & \cellcolor{red1}\textcolor{white}{202.3} & \cellcolor{red1}\textcolor{white}{45.6\%} \\
         RT-1-X & 195 & 43.0 & 58 & 13.3 & 22 & 0.7 & 4 & 0.0 & 279 & 57.0 & 20.4\% \\
         Octo-small & 8 & 0.3 & 15 & 3.3 & 21 & 1.0 & 10 & 0.3 & 54 & 4.9 & 9.1\% \\
         Octo-base & 0 & --- & 6 & 0.7 & 11 & 1.7 & 11 & 0.3 & 28 & 2.7 & 9.6\% \\
         OpenVLA-7b & 59 & 20.7 & 127 & 44 & 20 & 3.0 & 11 & 0.3 & \cellcolor{red4}217 & \cellcolor{red4}68.0 & \cellcolor{red4}31.3\% \\
         \hline
         \hline
         Tot. & 895 & 366.3 & 423 & 113.9 & 79 & 6.4 & 37 & 1.6 & 1434 & 488.2 & 34.0\%\\
         \hline
    \end{tabular}
    \begin{tablenotes}
        \item $\dagger$ Averaged results over three mutations.
    \end{tablenotes}
    \end{threeparttable}
    \vspace{-15pt}
\end{table}

Overall, we found that the VLA model's performance was greatly affected by the mutated camera poses. With mutated camera poses, the subject VLA models only passed 34.0\% of the test cases that could be passed with default camera poses. These results indicate that the current VLA models are very sensitive to the camera's extrinsic calibration results. When deploying VLA models, to achieve compatible performance, the developers need to carefully set up the camera poses, as in the robot demonstration data used for training. However, these may limit the generalizability of the VLA models. Future work may consider data augmentation to improve the VLA model's robustness under different camera settings to enhance generalizability.

\begin{finding}
    \label{finding:camera}
    Current VLA models are not robust against mutated camera poses, resulting in degraded performance when the visual input is captured from an angle that varies from the default one. When rotating the camera for a maximum angle of 5$^\circ$ and moving it for a maximum distance of 5cm, the performance dropped to 34.0\% of the performance on average with default camera poses. 
\end{finding}
\vspace{-1.5ex}

Among seven subject VLA models, the most robust one, RT-1-400k, succeeded in 45.6\% of the test cases with the mutated camera poses that were passed with the default camera poses. OpenVLA-7b also passed 31.3\% of the test cases. We noticed a significant performance gap between the two Octo models and other models, where both Octo models only passed less than 10\% of the test cases, while the other five models all passed more than 20\%. These results may have been largely attributed to the fact that Octo models used the smaller training dataset (\textasciitilde~65K robot demonstrations) compared with the other ones (130K~\textasciitilde~160K robot demonstrations). As a result, the Octo model's generalizability to visual inputs captured from different camera angles is significantly affected.

\begin{finding}
    \label{finding:camera_model}
    Among the seven subject VLA models, RT-1-400k was the most robust one against the perturbation of camera poses. Octo-series models performed significantly less robustly than the other models. This may be largely attributed to the fact that Octo models were trained with only half of the robot demonstration data compared with the other subject models.
\end{finding}
\vspace{-2ex}

\subsection{RQ5: \RQfive} 

To investigate this research question, we used an external object dataset, YCB~\cite{calli2015ycb}. YCB contains 56 objects that are not included in the Open-Embodiment-X dataset (the training/fine-tuning dataset of our seven subject VLA models). Similar to RQ1, we leveraged {\method} to generate 1,000 testing scenes for each of the four subject robotic manipulation tasks while sampling the target object and the \responseline{confounding objects} from the YCB dataset. We compared the performance of these subject VLA models with objects from the YCB dataset to the results in RQ1.

\begin{table}[t]
    \renewcommand{\arraystretch}{1}
    \centering
    \caption{\fboxsep\setlength{1pt} (\textbf{RQ5}) Performance of subject VLA models with seen/unseen objects. Each cell in the column \textit{Seen} and \textit{Unseen} represents the success rate of different VLA models in different tasks. \colorbox{red1}{\textcolor{white}{Darker}} the color, the larger the negative differences between the performance on seen and unseen objects.}
    \vspace{-10pt}
    \label{tab:unseen_objecs}
    \scriptsize
    \begin{threeparttable}
    \begin{tabular}{|l||rrr|rrr|rrr|rrr|}
         \hline
         \multirow{2}{*}{\textbf{VLA Models}} & \multicolumn{3}{c|}{\textbf{Task 1: Pick Up}} & \multicolumn{3}{c|}{\textbf{Task 2: Move Near}} & \multicolumn{3}{c|}{\textbf{Task 3: Put On}} & \multicolumn{3}{c|}{\textbf{Task 4: Put In}} \\
         \cline{2-13}
         & \multicolumn{1}{c}{Seen} & \multicolumn{1}{c}{Unseen} & \multicolumn{1}{c|}{Diff.$^\dagger$} & \multicolumn{1}{c}{Seen} & \multicolumn{1}{c}{Unseen} & \multicolumn{1}{c|}{Diff.$^\dagger$} & \multicolumn{1}{c}{Seen} & \multicolumn{1}{c}{Unseen} & \multicolumn{1}{c|}{Diff.$^\dagger$} & \multicolumn{1}{c}{Seen} & \multicolumn{1}{c}{Unseen} & \multicolumn{1}{c|}{Diff.$^\dagger$} \\
         \hline
         \hline
         RT-1-1k & 0.7\% & 1.4\% & \cellcolor{lightblue}+100.0\% & 1.4\% & 0.9\% & \cellcolor{lightred}-35.7\% & 0.0\% & 0.0\% & --- & 0.0\% & 0.0\% & --- \\
         RT-1-58k & 28.2\% & 5.0\% & \cellcolor{red1}\textcolor{white}{-82.3\%} & 10.9\% & 3.9\% & \cellcolor{red3}\textcolor{white}{-64.2\%} & 0.0\% & 0.0\% & --- & 0.0\% & 0.0\% & --- \\
         RT-1-400k & 34.4\% & 10.3\% & \cellcolor{red3}\textcolor{white}{-70.1\%} & 9.4\% & 2.9\% & \cellcolor{red3}\textcolor{white}{-69.2\%} & 0.5\% & 0.4\% & -\cellcolor{lightred}20.0\% & 0.1\% & 0.1\% & --- \\
         RT-1-X & 19.5\% & 3.0\% & \cellcolor{red1}\textcolor{white}{-84.6\%} & 5.8\% & 1.3\% & \cellcolor{red2}\textcolor{white}{-77.6\%} & 2.3\% & 0.6\% & \cellcolor{red2}\textcolor{white}{-73.9\%} & 0.4\% & 0.7\% & \cellcolor{lightblue}+75.0\% \\
         Octo-small & 0.8\% & 0.1\% & \cellcolor{red1}\textcolor{white}{-87.5\%} & 1.5\% & 1.5\% & --- & 2.2\% & 0.9\% & \cellcolor{red5}{-59.1\%} & 1.1\% & 0.8\% & \cellcolor{lightred}-27.3\% \\
         Octo-base & 0.0\% & 0.0\% & --- & 0.6\% & 0.4\% & \cellcolor{lightred}{-33.3\%} & 1.2\% & 0.2\% & \cellcolor{red1}\textcolor{white}{-83.3\%} & 1.1\% & 0.5\% & \cellcolor{red5}-54.5\% \\
         OpenVLA-7b & 5.9\% & 3.5\% &\cellcolor{lightred}-40.7\% & 12.7\% & 2.8\% & \cellcolor{red2}\textcolor{white}{-78.0\%} & 2.1\% & 0.9\% & \cellcolor{red5}{-57.1\%} & 1.1\% & 1.0\% & \cellcolor{lightred}-0.9\% \\
         \hline
         \hline
         Avg. & 12.8\% & 3.3\% & -74.2\% & 6.0\% & 2.0\% & -66.7\% & 1.2\% & 0.4\% & -66.7\% & 0.5\% & 0.4\% & -20.0\% \\
        \hline
    \end{tabular}
    \begin{tablenotes}
        \item $\dagger$ Diff. = (Unseen - Seen) / Seen
    \end{tablenotes}
    \end{threeparttable}
    \vspace{-12pt}
\end{table}

Table~\ref{tab:unseen_objecs} shows the performance of seven subject VLA models when manipulating seen and unseen objects. Overall, when manipulating unseen objects, the performance of the subject VLA models dropped significantly compared with manipulating seen objects. The average performance across different VLA models dropped by 74.2\%, 66.7\%, 66.7\%, and 20.0\% in Task 1, Task 2, Task 3, and Task 4, respectively. Notably, we also found that none of these VLA models completely failed when manipulating unseen objects (except those that also failed with seen objects). These results suggest that, while all VLA models exhibit certain possibilities of generalizing themselves to unseen objects, the performance is by far unreliable.

\begin{finding}
    \label{finding: unseen_objects}

    Current VLA models, though exhibit potential for generalization, are still unreliable for handling unseen objects. Our subject VLA models saw a performance drop ranging from 20.0\% to 74.2\% across four tasks when manipulating unseen objects compared with the seen objects.
    
\end{finding}
\vspace{-1.5ex}

We also examined each testing scene and the success of the VLA model at every individual step, similar to the investigation of RQ1 and RQ2, to understand how unseen objects might have impacted its performance. Specifically, since the model must succeed at each step sequentially to complete the testing scene, we define a metric called the \textbf{transfer rate} ($Tr$), which calculates the model’s ability to transfer success from one step to the next, as follows:

\vspace{-1.5ex}
\begin{equation}
    \label{eq:tr}
    \small
    Tr_n = \frac{Success\_rate_{n}}{Success\_rate_{n-1}},
\end{equation}
where $Success\_rate_{n}$ denotes the success rate of a task at step $n$. We define $Success\_rate_{0}=100.0\%$. 

\begin{figure}[t]
    \centering
    \includegraphics[width=0.95\linewidth]{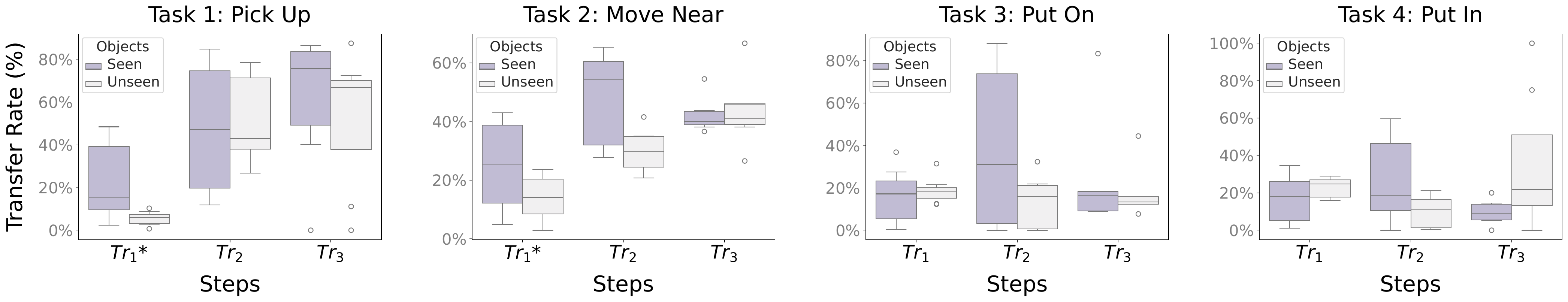}
    \vspace{-12pt}
    \caption{(\textbf{RQ5}) Transfer rate ($Tr$) of each task when manipulating with seen/unseen objects. Entries with * mean that their mean differences are statistically significant ($p<0.05$).}
    \label{fig:transfer_rate}
    \vspace{-15pt}
\end{figure}

We present the results in Fig.~\ref{fig:transfer_rate}. We also performed a paired $t$-test to examine the statistical differences between $Tr$ of seen and unseen objects. Overall, though $Tr_1$, $Tr_2$, and $Tr_3$ all decreased when manipulating with unseen objects instead of the seen ones, we only saw significant statistical differences on $Tr_1$ in Task 1 and Task 2 ($p = 0.011$ and $p=0.007$, Cohen's $d$ effect sizes are $1.34$ and $0.891$, respectively). These results reveal that the biggest challenge for current VLA models when dealing with unseen objects remains to be accurately locating the correct object to manipulate, suggesting that the VLA models may not be able to recognize unseen objects in many cases. We did not observe significant statistical differences on $Tr_1$ in Task 3 and Task 4. A plausible explanation is due to the small number of passed test cases (\textasciitilde~20 out of 1000) in both tasks.

\begin{finding}
    \label{finding: transfer_rate}

    Current VLA models struggled to recognize unseen objects, which was the primary cause of failure in most test scenes.
    
\end{finding}
\vspace{-2ex}

{\responseref{}

\subsection{RQ6: \RQsix}

To study this research question, for each task, we prompted GPT-4o to generate 10 mutations of the standard task instruction (Sec.~\ref{subsec:prompt}) that conveyed the same meanings. We only generated a limited number of mutations so that we could manually verify each mutation. Two of the authors manually validated and confirmed that all generated mutations were semantically equivalent to the corresponding standard task instruction. We then re-executed all test scenes generated in RQ1 (Sec.~\ref{subsec:rq1}) for each of the four tasks. For each test scene, we randomly applied one of the mutated task instructions. These mutated instructions can be found in our Git repository.

{\responseref{}
\begin{table}[t]
    \renewcommand{\arraystretch}{1}
    \centering
    \caption{\responseref{}\fboxsep\setlength{1pt} (\textbf{RQ6}) Performance of subject VLA models with standard (\textbf{Def.}) and mutated (\textbf{Mut.}) task instructions. Each cell represents the success rate of different VLA models in different tasks. \colorbox{red1}{\textcolor{white}{Darker}} the color, the larger the negative differences between the performance on standard and mutated task instructions.}
    \vspace{-10pt}
    \label{tab:task-mut}
    \scriptsize
    \begin{threeparttable}
    \begin{tabular}{|l||rrr|rrr|rrr|rrr|}
         \hline
         \multirow{2}{*}{\textbf{VLA Models}} & \multicolumn{3}{c|}{\textbf{Task 1: Pick Up}} & \multicolumn{3}{c|}{\textbf{Task 2: Move Near}} & \multicolumn{3}{c|}{\textbf{Task 3: Put On}} & \multicolumn{3}{c|}{\textbf{Task 4: Put In}} \\
         \cline{2-13}
         & \multicolumn{1}{c}{Def.} & \multicolumn{1}{c}{Mut.} & \multicolumn{1}{c|}{Diff.$^\dagger$} & \multicolumn{1}{c}{Def.} & \multicolumn{1}{c}{Mut.} & \multicolumn{1}{c|}{Diff.$^\dagger$} & \multicolumn{1}{c}{Def.} & \multicolumn{1}{c}{Mut.} & \multicolumn{1}{c|}{Diff.$^\dagger$} & \multicolumn{1}{c}{Def.} & \multicolumn{1}{c}{Mut.} & \multicolumn{1}{c|}{Diff.$^\dagger$} \\
         \hline
         \hline
         RT-1-1k & 0.7\% & 0.3\% & \cellcolor{red4}-57.1\% & 1.4\% & 0.9\% & \cellcolor{red5}-35.7\% & 0.0\% & 0.0\% & --- & 0.0\% & 0.0\% & --- \\
         RT-1-58k & 28.2\% & 18.8\% & \cellcolor{red5}-33.3\% & 10.9\% & 6.4\% & \cellcolor{red4}{-41.3\%} & 0.0\% & 0.0\% & --- & 0.0\% & 0.1\% & \cellcolor{lightblue}--- \\
         RT-1-400k & 34.4\% & 24.5\% & \cellcolor{red5}-28.7\% & 9.4\% & 8.8\% & \cellcolor{lightred}{-6.4\%} & 0.5\% & 0.4\% & -\cellcolor{lightred}20.0\% & 0.1\% & 0.1\% & --- \\
         RT-1-X & 19.5\% & 10.1\% & \cellcolor{red4}-48.2\% & 5.8\% & 5.6\% & \cellcolor{lightred}{-3.4\%} & 2.3\% & 1.5\% & \cellcolor{red5}{-34.8\%} & 0.4\% & 0.5\% & \cellcolor{lightblue}+25.0\% \\
         Octo-small & 0.8\% & 0.1\% & \cellcolor{red1}\textcolor{white}{-87.5\%} & 1.5\% & 1.5\% & --- & 2.2\% & 2.1\% & \cellcolor{lightred}{-4.5\%} & 1.1\% & 0.4\% & \cellcolor{red2}\textcolor{white}{-63.6\%} \\
         Octo-base & 0.0\% & 0.3\% & \cellcolor{lightblue}--- & 0.6\% & 0.4\% & \cellcolor{red5}{-33.3\%} & 1.2\% & 1.5\% & \cellcolor{lightblue}{+25.0\%} & 1.1\% & 0.5\% & \cellcolor{red4}-54.5\% \\
         OpenVLA-7b & 5.9\% & 6.4\% &\cellcolor{lightblue}+8.5\% & 12.7\% & 12.2\% & \cellcolor{lightred}{-0.4\%} & 2.1\% & 2.0\% & \cellcolor{lightred}{-4.8\%} & 1.1\% & 2.0\% & \cellcolor{lightblue}+81.8\% \\
         \hline
         \hline
         Avg. & 12.8\% & 8.6\% & -32.8\% & 6.0\% & 5.9\% & -1.7\% & 1.2\% & 1.1\% & -8.3\% & 0.5\% & 0.5\% & --- \\
        \hline
    \end{tabular}
    \begin{tablenotes}
        \item $\dagger$ Diff. = (Mut. - Def.) / Def.
    \end{tablenotes}
    \end{threeparttable}
    \vspace{-15pt}
\end{table}
}

Table~\ref{tab:task-mut} presents the performance of seven subject VLA models when instructed with both the standard and mutated task instructions. When using the mutated task instructions, the average performance across the seven VLA models dropped by 32.8\%, 1.7\%, and 8.3\% on Task 1, Task 2, and Task 3, respectively, while the performance difference on Task 4 was negligible.

We also found that larger models, such as OpenVLA-7b, were more robust against task instruction mutations. Specifically, OpenVLA-7b performed even better with the mutated task instructions on Task 1 and Task 4. For Task 2 and Task 3, the performance was only marginally affected. This robustness may be attributed to OpenVLA-7b’s use of Llama 2 as its language model, which potentially enhanced its language comprehension capabilities.

For instance, a mutated task instruction for Task 2 could be: {\small \texttt{place [OBJECT A] near [OBJECT B]}}. When prompted with this mutated instruction, OpenVLA-7b successfully manipulated 17.6\% of the test scenes. In contrast, none of the other VLA models succeeded in more than 5\% of the test scenes under the same instruction.

\begin{finding}
    \responseline{Current VLA models showed limited robustness against task instruction mutations. However, VLA models incorporating a large language model (e.g., OpenVLA-7b uses Llama 2-7b as the language model) are much more robust.}
\end{finding}
\vspace{-2ex}

}

\section{Discussion}
\label{sec:discussion}

Our study reveals several key implications that can lead to the development of better and more reliable VLA models for robotic manipulation. In this section, we discuss these implications and explore future research opportunities.

\smalltitle{VLA models for robotic manipulation---too good to be true for now.} While VLA models have the potential to revolutionize robotic manipulation, our experiment results in RQ1 and RQ2 reveal that the current models are still \textbf{unreliable} for common robotic manipulation tasks at the time of this study. Deploying VLA models in high-stakes and safety-critical applications remains impractical.

Our detailed analysis suggests that the deficiency mainly comes from the lack of capabilities in precisely interpreting complex task requirements and accurately localizing the correct target objects to manipulate. We believe there is a huge space for improvement in addressing these issues. One potential solution is to scale up the model size. Currently, even the largest model in our study, OpenVLA-7b, has only 7 billion parameters, which is significantly smaller than state-of-the-art (SOTA) models in other domains. For instance, one of the largest open-source LLMs, Llama 3.1, has 405 billion parameters. The closed-source LLMs such as GPT-4 and Claude-3.5 typically have even higher numbers of model parameters. By scaling up the model, we may also observe \textit{emergent capabilities} for VLA models~\cite{brohan2023rt}.

In addition to scaling up the size of VLA models, exploring more effective prompting strategies could be another promising direction. Previous studies have demonstrated that improved prompting can significantly enhance the performance of LLMs across various tasks, such as solving mathematical problems~\cite{wei2022chain, yao2024tree} and code generation~\cite{zhang2023self, chen2023teaching}. Our Finding~\ref{findings:complex_command} suggests that current VLA models struggle to decompose complex task instructions into multiple steps of action. To fix this, one may consider prompting the VLA model step by step.
This also aligns with one of the popular prompting strategies in the field of LLMs -- \textit{chain-of-thought} prompting. While our study has only explored the most basic prompts, future research should explore more advanced prompting techniques to evaluate their impact on VLA model performance.

\begin{wraptable}{r}{0.55\linewidth}
\vspace{-12pt}
\renewcommand{\arraystretch}{1}
\caption{\responseref{} Different prompts for Task 4.}
\vspace{-10pt}
\label{tab:improve_prompt}
\scriptsize
\centering
\begin{tabular}{|l|r|r|r||r|r|r|}
\hline
\multirow{2}{*}{\textbf{Model}} & \multicolumn{3}{c||}{\textbf{Original Prompt}} & \multicolumn{3}{c|}{\textbf{New Prompt}} \\
\cline{2-7}
& Grasp & Move & Success & Grasp & Move & Success \\
\hline
\hline
RT-1-1k & 1.1\% & 0.0\% & 0.0\% & \cellcolor{red5}1.4\% & 0.0\% & 0.0\% \\
RT-1-58k & 1.3\% & 0.0\% & 0.0\% & \cellcolor{red5}4.2\% & \cellcolor{red5}0.6\% & 0.0\% \\
RT-1-400k & 8.9\% & 0.1\% & 0.1\% & 6.0\% & 0.1\% & 0.0\% \\
RT-1-X & 17.8\% & 0.7\% & 0.4\% & \cellcolor{red5}28.0\% & \cellcolor{red5}7.8\% & 0.4\% \\
Octo-small & 34.6\% & 1.1\% & 1.1\% & 30.0\% & \cellcolor{red5}17.4\% & 0.0\% \\
Octo-base & 32.5\% & 1.1\% & 1.1\% & 22.1\% & \cellcolor{red5}12.9\% & 0.2\% \\
OpenVLA & 19.9\% & 1.1\% & 1.1\% & 14.1\% & \cellcolor{red5}4.4\% & 0.1\% \\ 
\hline
\end{tabular}
\vspace{-10pt}
\end{wraptable}\responseline{To demonstrate the potential of improving VLA models’ performance through prompting, we crafted a new prompt for Task 4 by separating the task into two steps: {\small \texttt{pick up [object name] and then put [object name] into [object name]}}. We re-ran the 1000 test scenes generated in RQ1 (Sec.~\ref{subsec:rq1}) for Task 4 using this new prompt and compared the performance with the original prompt. Our results in Table~\ref{tab:improve_prompt} show that with this new prompt, the success rate of grasping the correct object increased in three out of seven VLA models, while the success rate of moving the correct object improved in five out of seven models. The average success rate of moving the correct object increased from 0.6\% to 6.2\%. Although the overall task success rate did not increase, these results suggest that exploring prompting strategies for VLA models could be a promising research direction.
}

Furthermore, one can also introduce multi-agent systems to split the robotic manipulation tasks among multiple VLA agents, which is a strategy widely used in other domains~\cite{xia2024agentless, zhang2024autocoderover, yang2024swe}.

\smalltitle{Addressing the robustness challenges.} Our study results from RQ3 and RQ4 reveal that current VLA models lack robustness against several external factors, such as lighting conditions and camera poses. We also found that models with large-scale pre-training or those trained with larger datasets of robot demonstration data exhibit greater robustness compared to others. This highlights a promising research direction: enriching the robot demonstration data. Since manually collecting real-world robot demonstration data requires significant labeling efforts, researchers may consider leveraging data augmentation techniques~\cite{alomar2023data, bharadhwaj2024roboagent} or employing \textit{sim2real} translation~\cite{zhao2020sim, hofer2021sim2real} to scale up training data by utilizing simulation environments. \responseline{For instance, in the future, one could leverage a well-designed traditional controller to guide the robot in solving a test scene. The robot’s demonstration could then serve as data for re-training or fine-tuning.}

Moreover, our Finding~\ref{finding:increase_lighting}, Finding~\ref{finding:decrease_lighting}, and Finding~\ref{finding:camera} show that when external factors like lighting and camera angles deviate significantly from default settings, VLA model performance drops accordingly. This suggests that existing robot demonstration datasets may lack diversity, especially regarding variations in external conditions. In future work, when collecting datasets for training/fine-tuning factors, practitioners should take these environmental factors into account, which may potentially lead to the training of more robust VLA models.

\smalltitle{Assessing the capabilities of VLA models.} Our Finding~\ref{finding: unseen_objects} and Finding~\ref{finding: transfer_rate} (RQ5) suggest that current VLA models struggled to perform tasks with unseen objects. In practice, it is unrealistic to expect VLA models to successfully perform manipulation for every possible task scenario. Thus, in parallel to the development of more powerful and robust VLA models, it is also important to design novel techniques to comprehensively assess the capabilities of VLA models and derive proper guidelines about the use of VLA models. To address these, potential solutions include offline benchmarking and online risk assessment.

In terms of offline benchmarking, our 18,604 generated test scenes across four tasks can serve as one of the early benchmarks for VLA models. In future work, practitioners may focus on expanding this benchmark to include a broader range of robotic manipulation tasks and diverse robot settings, covering various object types, environmental factors, and task complexities. In terms of online risk assessment, one may consider adopting SOTA techniques for risk assessment in other domains, such as uncertainty estimation~\cite{huang2023look, mukherjee2020uncertainty} and safety monitoring~\cite{xie2024online}. While general techniques may be directly used in the context of robotic manipulation with VLA models, it is unclear to what extent they can help with assessing the quality and reliability of decisions made by VLA models. 

\smalltitle{Towards efficient testing for VLA models.} Although our proposed {\method} successfully identified numerous failed test scenes across our subject VLA models, it also incurred significant time overheads. Specifically, we relied on a \textit{random sampler} for sampling target objects and \responseline{confounding objects} in the testing scenes, which may not have been the most efficient approach. Future research could focus on optimizing the $pose_sampler$ in Algorithm~\ref{alg:test_generation} to strategically assign critical positions and orientations for objects and \responseline{confounding objects}. For instance, one can consider metamorphic-based methods~\cite{wang2020metamorphic, yu2022automated, gao2024multitest} or search-based methods~\cite{zhou2023specification, cheng2023behavexplor} to efficiently generate test scenes for VLA models. Meanwhile, researchers may also work on test prioritization~\cite{li2024prioritizing, dang2024test, feng2020deepgini} or test selection~\cite{huang2024active, hu2023aries, gao2022adaptive, aghababaeyan2024deepgd} towards efficient testing for VLA models. 

Nevertheless, as VLA models continue to evolve, we argue that testing strategies must also evolve in parallel. The complexity and capabilities of these models are likely to increase, requiring more sophisticated and adaptive testing frameworks that can keep pace with advancements.


\section{Related Work}
\label{sec:related_work}


\smalltitle{Foundation Models for Robotics.}
A large body of research has been done on the use of foundation models for robotics. Based on the data modality, recent work can be roughly categorized into two groups: (1) those that use LLMs for robotics and (2) those that use multi-modality foundation models (e.g., VLMs) for robotics.
One of the early attempts in using LLMs for robotics is to leverage LLMs as a reward designer for Deep Reinforcement Learning-based robotic manipulation~\cite{song2023self}. 
This work proposes a self-refined framework to let the LLM automatically generate and refine reward functions used to train policies for robotic control. 
Alternatively, Zhou~\etal~\cite{zhou2024isr} adopts LLMs to convert natural language input into a Planning Domain Definition Language formulation before outputting action plan sequences.


Different from the use of LLMs, multi-modality foundation models such as VLMs enable the possibility of advancing AI-enabled robotics in several novel tasks, e.g., for robotic manipulation~\cite{brohan2022rt, brohan2023rt, team2024octo}, visual question answering with robots~\cite{du2023vision}, and visual state representations~\cite{nair2022r3m}. Our work is most closely related to those using vision-language-action models, a specialized category of VLMs, for robotic manipulation~\cite{brohan2023rt, team2024octo, kim2024openvla, li2024llara, brohan2022rt, li2023vision, stone2023open}. One of the pioneer works in VLA models for robotic manipulation is RT-1~\cite{brohan2022rt}, which uses a combination of a FiLM EfficientNet and a transformer to learn control policies from 130k real-world robot demonstrations. Since the release of the Open X-Embodiment dataset~\cite{padalkar2023open}, a series of VLA models have been proposed by either training or fine-tuning on Open X-Embodiment~\cite{brohan2023rt, team2024octo, kim2024openvla, li2024llara}. Our work is parallel to these works. We propose a general testing framework to test VLA models and present an empirical study to comprehensively assess the performance of VLA models.

\smalltitle{Testing Cyber-Physical Systems.}
Quality assurance of traditional CPS and AI-driven CPS is a vital topic in software engineering, as it ensures the safety and trustworthiness of such complex systems in safety-critical domains.
Substantial efforts have been made by researchers and industrial practitioners to safeguard the quality of CPS from various perspectives, such as testing~\cite{lee2023fuzzing, menghi2020approximation, chen2020active, ayerdi2021generating, zhang2022falsifai}, analysis~\cite{stocco2022confidence, biagiola2024testing, yohanandhan2020cyber} and repairing~\cite{song2023mathtt, wang2024mortar, guo2020cps}.
Considering robotic manipulation as a widely deployed representative CPS, our work is most related to the testing of CPS.

In particular, Lee~\etal~\cite{lee2023fuzzing} propose MOTIF, a gray-box fuzzing-based approach to generate test data for software deployed in CPS. 
This method monitors the coverage achieved by the vanilla and the mutated functions, thereby exploring the behavior space of the software under test.
Differing from the existing testing solutions, MOTIF specifically targets the testing challenges of C and C++ languages that are widely used in CPS contexts. 
In terms of the black-box approach, Menghi~\etal~\cite{menghi2020approximation} introduce ARIsTEO, a search-based testing method with a loop of approximation and refinement to identify requirement violations of CPS models, and Chen~\etal~\cite{chen2020active} propose an active fuzzing approach to find test suites for packet-level CPS network attacks.
Moving forward with CPS with an AI module embedded, Zhang~\etal~\cite{zhang2022falsifai} design a series of time-aware coverage criteria for DNN controllers in CPS and develop a falsification framework, FalsifAI, to find system defects by leveraging the coverage information from the proposed criteria.
It is worth noting that the aforementioned works are not well applicable to VLA models, which is attributable to the multimodality characteristic, the autoregressive generation mechanism, and the large-scale model size.
In contrast, our study initiates a testing framework tailored for VLA models in the context of different robotic manipulation tasks.
{\method} encompasses four multimodality-aware testing operators to comprehensively assess the VLA models from different perspectives.  



\smalltitle{Benchmarking and Testing Foundation Models.}
After the remarkable success of foundation models, an important research direction is to explore and understand the boundaries of their capabilities, which can provide a basis for further safeguarding and enhancement. Based on this motivation, numerous studies have focused on evaluating large language models in the text domain across various properties, such as correctness~\cite{liang2023holistic}, factuality~\cite{lin2021truthfulqa, dziri2022faithdial}, robustness~\cite{zhu2023promptbench, wang2023decodingtrust}, fairness~\cite{sun2024trustllm}, and privacy~\cite{hamid2023genaipabench}. These studies offer a wealth of resources for the general performance assessment of LLMs. On the other hand, researchers have also made significant efforts to evaluate specific capabilities of LLMs, such as code-related abilities. These efforts include validating the correctness of generated code~\cite{cassano2023multipl, liu2024your, yu2024codereval}, examining whether LLMs can address real-world GitHub issues~\cite{jimenez2023swe}, and analyzing error patterns of LLM-generated code~\cite{pan2024lost, wang2025towards}. 

While the aforementioned studies provide valuable insights, most of them focus on performing analysis on static benchmarks. Orthogonal to these related works, there are also several attempts at dynamically generating test cases to evaluate one or more key properties of models. For example, existing studies use rule-based methods to generate test cases that measure bias~\cite{wan2023biasasker} or linguistic capabilities~\cite{lee2024automated}, employ metamorphic testing for evaluating language translation~\cite{sun2020automatic, he2020structure}, and utilize mutation-based frameworks to assess robustness~\cite{xiao2023leap}. Our work differs from these previous approaches in two main ways: (1) our framework centers on VLA, a novel architecture that integrates multi-modal inputs and generates complex control commands, distinguishing it from language-focused foundation models; (2) our framework generates test cases that account for complex interactions in robotic manipulation within a 3D environment, presenting a new challenge compared to prior text-only testing frameworks.




\section{Threats to Validity}
\label{sec:threats}

We discuss the threats to the validity of our empirical study results, as well as the mitigating factors.

\smalltitle{Internal Validity.} One potential threat comes from the randomness of our experiments. To mitigate this, we generated 18,604 testing scenes in our empirical study. Overall, our empirical study took over \responseline{580} GPU hours to finish. While we only executed each testing scene once per VLA model, we believe this has limited impact on our findings as our goal is to evaluate the VLA model's overall performance on each task instead of the success of one specific testing scene. 

Another potential threat to internal validity is the choice of prompt templates. \responseline{To mitigate this, we followed the previous works~\cite{team2024octo, kim2024openvla, brohan2022rt, brohan2023rt} to use the standard prompt template when evaluating VLA models for robotic manipulation in RQ1~\textasciitilde~RQ5. In RQ6, we evaluated the VLA models' performance with the mutated prompt templates. However, the prompt templates mutated by GPT-4o may not carry the same meanings as the original ones. To mitigate the threat, two of the authors manually validated that the mutated prompts were semantically equivalent to the corresponding standard prompts.} In addition, we discuss the impact of prompts and the potential of improving VLA models with better prompting in Sec.~\ref{sec:discussion}.

\smalltitle{External Validity.} Threats to external validity include whether our study results can generalize to different experimental settings, such as different robotic manipulation tasks and different VLA models. To mitigate these threats, we select the four most popular robotic manipulation tasks according to the Open X-Embodiment dataset~\cite{padalkar2023open}. In terms of the VLA models, we included the SOTA publicly available VLA models by the time we conducted this study. Given that the VLA model for robotic manipulation is a fast-developing research area, we also plan to extend our evaluation to other SOTA VLA models, e.g., RT-2~\cite{brohan2023rt}, once they are made available. 

Finally, we have only implemented {\method} and evaluated the VLA models within one simulation environment, Maniskill2~\cite{gu2023maniskill2}. To reduce the impact of distribution shifts between simulation and real robot execution, we chose an improved version of Maniskill2~\cite{li2024evaluating}, which provides \textit{visual matching} to render realistic images as the visual input for VLA models. 

\smalltitle{Construct Validity.} When designing our testing framework, {\tool}, we have only considered a limited number of testing operators. To combat the threat, we included testing operators covering different aspects, e.g., task difficulties (i.e., number of \responseline{confounding objects}) and environmental factors (i.e., the lighting intensity and camera poses). In future work, one may continue to improve {\method} by adding more testing operators, such as the number of lighting sources, the intrinsic parameters of cameras, and the resolution of cameras, for a more comprehensive evaluation of VLA models. {\responseref{} Another threat to construct validity is the coverage metric we used in RQ1. Although our trajectory coverage metric may not be the best metric for measuring test sufficiency, we believe it is a pragmatic choice since there is no well-established coverage metric for either VLA models or robotic manipulation. In future work, one may consider designing a new coverage metric given the special characteristics of robotic manipulation with VLA models.}

\section{Conclusion}

In this paper, we propose {\method}, one of the early testing frameworks for testing VLA models for robotic manipulation. {\method} is a generation-based fuzzing framework based on ten operators covering different perspectives in the test scene of robotic manipulation. Upon implementing {\method} with Maniskill2 simulation environments, we further conducted a large-scale empirical study to assess the performance and robustness of seven popular VLA models across four robotic manipulation tasks. We generated 18,604 testing scenes, conducted more than \responseline{580} GPU hours of simulation, and performed a detailed analysis of the challenges and limitations faced by the current VLA models. At the end of the paper, we discuss the implications of our study, shedding light on several future research directions to improve the quality and reliability of VLA models.

\section{Data Availability}

Our artifacts, including the replication packages and the generated testing scenes, are available on a Git repository: \href{https://github.com/ma-labo/VLATest}{https://github.com/ma-labo/VLATest}.

\begin{acks}
This work was supported in part by the Canada CIFAR AI Chairs Program, Natural Sciences and Engineering Research Council of Canada, JST CRONOS Grant (No. JPMJCS24K8), JSPS KAKENHI Grant (No. JP21H04877, No. JP23H03372, and No. JP24K02920), and the Autoware Foundation.
\end{acks}

\bibliographystyle{ACM-Reference-Format}
\bibliography{references}


\begin{thebibliography}{99}


\ifx \showCODEN    \undefined \def \showCODEN     #1{\unskip}     \fi
\ifx \showISBNx    \undefined \def \showISBNx     #1{\unskip}     \fi
\ifx \showISBNxiii \undefined \def \showISBNxiii  #1{\unskip}     \fi
\ifx \showISSN     \undefined \def \showISSN      #1{\unskip}     \fi
\ifx \showLCCN     \undefined \def \showLCCN      #1{\unskip}     \fi
\ifx \shownote     \undefined \def \shownote      #1{#1}          \fi
\ifx \showarticletitle \undefined \def \showarticletitle #1{#1}   \fi
\ifx \showURL      \undefined \def \showURL       {\relax}        \fi
\providecommand\bibfield[2]{#2}
\providecommand\bibinfo[2]{#2}
\providecommand\natexlab[1]{#1}
\providecommand\showeprint[2][]{arXiv:#2}

\bibitem[Aghababaeyan et~al\mbox{.}(2024)]%
        {aghababaeyan2024deepgd}
\bibfield{author}{\bibinfo{person}{Zohreh Aghababaeyan}, \bibinfo{person}{Manel Abdellatif}, \bibinfo{person}{Mahboubeh Dadkhah}, {and} \bibinfo{person}{Lionel Briand}.} \bibinfo{year}{2024}\natexlab{}.
\newblock \showarticletitle{DeepGD: A Multi-Objective Black-Box Test Selection Approach for Deep Neural Networks}.
\newblock \bibinfo{journal}{\emph{ACM Transactions on Software Engineering and Methodology}} \bibinfo{volume}{33}, \bibinfo{number}{6} (\bibinfo{year}{2024}), \bibinfo{pages}{1--29}.
\newblock


\bibitem[Alomar et~al\mbox{.}(2023)]%
        {alomar2023data}
\bibfield{author}{\bibinfo{person}{Khaled Alomar}, \bibinfo{person}{Halil~Ibrahim Aysel}, {and} \bibinfo{person}{Xiaohao Cai}.} \bibinfo{year}{2023}\natexlab{}.
\newblock \showarticletitle{Data Augmentation in Classification and Segmentation: A Survey and New Strategies}.
\newblock \bibinfo{journal}{\emph{Journal of Imaging}} \bibinfo{volume}{9}, \bibinfo{number}{2} (\bibinfo{year}{2023}), \bibinfo{pages}{46}.
\newblock


\bibitem[Ayerdi et~al\mbox{.}(2021)]%
        {ayerdi2021generating}
\bibfield{author}{\bibinfo{person}{Jon Ayerdi}, \bibinfo{person}{Valerio Terragni}, {et~al\mbox{.}}} \bibinfo{year}{2021}\natexlab{}.
\newblock \showarticletitle{Generating metamorphic relations for cyber-physical systems with genetic programming: an industrial case study}. In \bibinfo{booktitle}{\emph{Proceedings of the 29th ACM Joint Meeting on European Software Engineering Conference and Symposium on the Foundations of Software Engineering}}. \bibinfo{pages}{1264--1274}.
\newblock


\bibitem[Bai et~al\mbox{.}(2022)]%
        {bai2022training}
\bibfield{author}{\bibinfo{person}{Yuntao Bai}, \bibinfo{person}{Andy Jones}, \bibinfo{person}{Kamal Ndousse}, {et~al\mbox{.}}} \bibinfo{year}{2022}\natexlab{}.
\newblock \showarticletitle{Training a Helpful and Harmless Assistant with Reinforcement Learning from Human Feedback}.
\newblock \bibinfo{journal}{\emph{arXiv preprint arXiv:2204.05862}} (\bibinfo{year}{2022}).
\newblock


\bibitem[Bharadhwaj et~al\mbox{.}(2024)]%
        {bharadhwaj2024roboagent}
\bibfield{author}{\bibinfo{person}{Homanga Bharadhwaj}, \bibinfo{person}{Jay Vakil}, \bibinfo{person}{Mohit Sharma}, \bibinfo{person}{Abhinav Gupta}, \bibinfo{person}{Shubham Tulsiani}, {and} \bibinfo{person}{Vikash Kumar}.} \bibinfo{year}{2024}\natexlab{}.
\newblock \showarticletitle{RoboAgent: Generalization and Efficiency in Robot Manipulation via Semantic Augmentations and Action Chunking}. In \bibinfo{booktitle}{\emph{2024 IEEE International Conference on Robotics and Automation (ICRA)}}. IEEE, \bibinfo{pages}{4788--4795}.
\newblock


\bibitem[Biagiola and Tonella(2024)]%
        {biagiola2024testing}
\bibfield{author}{\bibinfo{person}{Matteo Biagiola} {and} \bibinfo{person}{Paolo Tonella}.} \bibinfo{year}{2024}\natexlab{}.
\newblock \showarticletitle{Testing of Deep Reinforcement Learning Agents with Surrogate Models}.
\newblock \bibinfo{journal}{\emph{ACM Transactions on Software Engineering and Methodology}} \bibinfo{volume}{33}, \bibinfo{number}{3} (\bibinfo{year}{2024}), \bibinfo{pages}{1--33}.
\newblock


\bibitem[Brohan et~al\mbox{.}(2022)]%
        {brohan2022rt}
\bibfield{author}{\bibinfo{person}{Anthony Brohan}, \bibinfo{person}{Noah Brown}, \bibinfo{person}{Justice Carbajal}, {et~al\mbox{.}}} \bibinfo{year}{2022}\natexlab{}.
\newblock \showarticletitle{RT-1: Robotics Transformer for Real-World Control at Scale}.
\newblock \bibinfo{journal}{\emph{arXiv preprint arXiv:2212.06817}} (\bibinfo{year}{2022}).
\newblock


\bibitem[Brohan et~al\mbox{.}(2023)]%
        {brohan2023rt}
\bibfield{author}{\bibinfo{person}{Anthony Brohan}, \bibinfo{person}{Noah Brown}, \bibinfo{person}{Justice Carbajal}, {et~al\mbox{.}}} \bibinfo{year}{2023}\natexlab{}.
\newblock \showarticletitle{RT-2: Vision-Language-Action Models Transfer Web Knowledge to Robotic Control}.
\newblock \bibinfo{journal}{\emph{arXiv preprint arXiv:2307.15818}} (\bibinfo{year}{2023}).
\newblock


\bibitem[Calli et~al\mbox{.}(2015)]%
        {calli2015ycb}
\bibfield{author}{\bibinfo{person}{Berk Calli}, \bibinfo{person}{Arjun Singh}, \bibinfo{person}{Aaron Walsman}, \bibinfo{person}{Siddhartha Srinivasa}, \bibinfo{person}{Pieter Abbeel}, {and} \bibinfo{person}{Aaron~M Dollar}.} \bibinfo{year}{2015}\natexlab{}.
\newblock \showarticletitle{The YCB object and Model set: Towards common benchmarks for manipulation research}. In \bibinfo{booktitle}{\emph{2015 international conference on advanced robotics (ICAR)}}. IEEE, \bibinfo{pages}{510--517}.
\newblock


\bibitem[Cassano et~al\mbox{.}(2023)]%
        {cassano2023multipl}
\bibfield{author}{\bibinfo{person}{Federico Cassano}, \bibinfo{person}{John Gouwar}, \bibinfo{person}{Daniel Nguyen}, {et~al\mbox{.}}} \bibinfo{year}{2023}\natexlab{}.
\newblock \showarticletitle{MultiPL-E: A Scalable and Extensible Approach to Benchmarking Neural Code Generation}.
\newblock \bibinfo{journal}{\emph{IEEE Transactions on Software Engineering}} \bibinfo{volume}{49}, \bibinfo{number}{7} (\bibinfo{year}{2023}), \bibinfo{pages}{3675--3691}.
\newblock


\bibitem[Chen et~al\mbox{.}(2024)]%
        {chen2023teaching}
\bibfield{author}{\bibinfo{person}{Xinyun Chen}, \bibinfo{person}{Maxwell Lin}, \bibinfo{person}{Nathanael Sch{\"a}rli}, {and} \bibinfo{person}{Denny Zhou}.} \bibinfo{year}{2024}\natexlab{}.
\newblock \showarticletitle{Teaching Large Language Models to Self-Debug}. In \bibinfo{booktitle}{\emph{The Twelfth International Conference on Learning Representations}}.
\newblock
\urldef\tempurl%
\url{https://openreview.net/forum?id=KuPixIqPiq}
\showURL{%
\tempurl}


\bibitem[Chen et~al\mbox{.}(2020)]%
        {chen2020active}
\bibfield{author}{\bibinfo{person}{Yuqi Chen}, \bibinfo{person}{Bohan Xuan}, \bibinfo{person}{Christopher~M Poskitt}, \bibinfo{person}{Jun Sun}, {and} \bibinfo{person}{Fan Zhang}.} \bibinfo{year}{2020}\natexlab{}.
\newblock \showarticletitle{Active Fuzzing for Testing and Securing Cyber-Physical Systems}. In \bibinfo{booktitle}{\emph{Proceedings of the 29th ACM SIGSOFT International Symposium on Software Testing and Analysis}}. \bibinfo{pages}{14--26}.
\newblock


\bibitem[Cheng et~al\mbox{.}(2023)]%
        {cheng2023behavexplor}
\bibfield{author}{\bibinfo{person}{Mingfei Cheng}, \bibinfo{person}{Yuan Zhou}, {and} \bibinfo{person}{Xiaofei Xie}.} \bibinfo{year}{2023}\natexlab{}.
\newblock \showarticletitle{BehAVExplor: Behavior Diversity Guided Testing for Autonomous Driving Systems}. In \bibinfo{booktitle}{\emph{Proceedings of the 32nd ACM SIGSOFT International Symposium on Software Testing and Analysis}}. \bibinfo{pages}{488--500}.
\newblock


\bibitem[Dang et~al\mbox{.}(2024)]%
        {dang2024test}
\bibfield{author}{\bibinfo{person}{Xueqi Dang}, \bibinfo{person}{Yinghua Li}, \bibinfo{person}{Mike Papadakis}, \bibinfo{person}{Jacques Klein}, \bibinfo{person}{Tegawend{\'e}~F Bissyand{\'e}}, {and} \bibinfo{person}{Yves Le~Traon}.} \bibinfo{year}{2024}\natexlab{}.
\newblock \showarticletitle{Test Input Prioritization for Machine Learning Classifiers}.
\newblock \bibinfo{journal}{\emph{IEEE Transactions on Software Engineering}} (\bibinfo{year}{2024}).
\newblock


\bibitem[Dhaliwal(2020)]%
        {dhaliwal2020rise}
\bibfield{author}{\bibinfo{person}{Amandeep Dhaliwal}.} \bibinfo{year}{2020}\natexlab{}.
\newblock \showarticletitle{The Rise of Automation and Robotics in Warehouse Management}.
\newblock In \bibinfo{booktitle}{\emph{Transforming Management Using Artificial Intelligence Techniques}}. \bibinfo{publisher}{CRC Press}, \bibinfo{pages}{63--72}.
\newblock


\bibitem[Ding et~al\mbox{.}(2023)]%
        {ding2023task}
\bibfield{author}{\bibinfo{person}{Yan Ding}, \bibinfo{person}{Xiaohan Zhang}, \bibinfo{person}{Chris Paxton}, {and} \bibinfo{person}{Shiqi Zhang}.} \bibinfo{year}{2023}\natexlab{}.
\newblock \showarticletitle{Task and Motion Planning with Large Language Models for Object Rearrangement}. In \bibinfo{booktitle}{\emph{2023 IEEE/RSJ International Conference on Intelligent Robots and Systems (IROS)}}. IEEE, \bibinfo{pages}{2086--2092}.
\newblock


\bibitem[Dosovitskiy et~al\mbox{.}(2020)]%
        {dosovitskiy2020image}
\bibfield{author}{\bibinfo{person}{Alexey Dosovitskiy}, \bibinfo{person}{Lucas Beyer}, \bibinfo{person}{Alexander Kolesnikov}, \bibinfo{person}{Dirk Weissenborn}, \bibinfo{person}{Xiaohua Zhai}, \bibinfo{person}{Thomas Unterthiner}, \bibinfo{person}{Mostafa Dehghani}, \bibinfo{person}{Matthias Minderer}, \bibinfo{person}{Georg Heigold}, \bibinfo{person}{Sylvain Gelly}, {et~al\mbox{.}}} \bibinfo{year}{2020}\natexlab{}.
\newblock \showarticletitle{An Image is Worth 16x16 Words: Transformers for Image Recognition at Scale}. In \bibinfo{booktitle}{\emph{International Conference on Learning Representations}}.
\newblock


\bibitem[Driess et~al\mbox{.}(2023)]%
        {driess2023palm}
\bibfield{author}{\bibinfo{person}{Danny Driess}, \bibinfo{person}{Fei Xia}, \bibinfo{person}{Mehdi S.~M. Sajjadi}, {et~al\mbox{.}}} \bibinfo{year}{2023}\natexlab{}.
\newblock \showarticletitle{{P}a{LM}-E: An Embodied Multimodal Language Model}. In \bibinfo{booktitle}{\emph{Proceedings of the 40th International Conference on Machine Learning}} \emph{(\bibinfo{series}{Proceedings of Machine Learning Research}, Vol.~\bibinfo{volume}{202})}. \bibinfo{publisher}{PMLR}, \bibinfo{pages}{8469--8488}.
\newblock


\bibitem[Du et~al\mbox{.}(2023)]%
        {du2023vision}
\bibfield{author}{\bibinfo{person}{Yuqing Du}, \bibinfo{person}{Ksenia Konyushkova}, \bibinfo{person}{Misha Denil}, \bibinfo{person}{Akhil Raju}, \bibinfo{person}{Jessica Landon}, \bibinfo{person}{Felix Hill}, \bibinfo{person}{Nando de Freitas}, {and} \bibinfo{person}{Serkan Cabi}.} \bibinfo{year}{2023}\natexlab{}.
\newblock \showarticletitle{Vision-Language Models as Success Detectors}. In \bibinfo{booktitle}{\emph{Proceedings of The 2nd Conference on Lifelong Learning Agents}} \emph{(\bibinfo{series}{Proceedings of Machine Learning Research}, Vol.~\bibinfo{volume}{232})}. \bibinfo{publisher}{PMLR}, \bibinfo{pages}{120--136}.
\newblock


\bibitem[Dzedzickis et~al\mbox{.}(2021)]%
        {dzedzickis2021advanced}
\bibfield{author}{\bibinfo{person}{Andrius Dzedzickis}, \bibinfo{person}{Jurga Suba{\v{c}}i{\=u}t{\.e}-{\v{Z}}emaitien{\.e}}, \bibinfo{person}{Ernestas {\v{S}}utinys}, \bibinfo{person}{Urt{\.e} Samukait{\.e}-Bubnien{\.e}}, {and} \bibinfo{person}{Vytautas Bu{\v{c}}inskas}.} \bibinfo{year}{2021}\natexlab{}.
\newblock \showarticletitle{Advanced applications of industrial robotics: New trends and possibilities}.
\newblock \bibinfo{journal}{\emph{Applied Sciences}} \bibinfo{volume}{12}, \bibinfo{number}{1} (\bibinfo{year}{2021}), \bibinfo{pages}{135}.
\newblock


\bibitem[Dziri et~al\mbox{.}(2022)]%
        {dziri2022faithdial}
\bibfield{author}{\bibinfo{person}{Nouha Dziri}, \bibinfo{person}{Ehsan Kamalloo}, \bibinfo{person}{Sivan Milton}, \bibinfo{person}{Osmar Zaiane}, \bibinfo{person}{Mo Yu}, \bibinfo{person}{Edoardo~M Ponti}, {and} \bibinfo{person}{Siva Reddy}.} \bibinfo{year}{2022}\natexlab{}.
\newblock \showarticletitle{FaithDial: A Faithful Benchmark for Information-Seeking Dialogue}.
\newblock \bibinfo{journal}{\emph{Transactions of the Association for Computational Linguistics}}  \bibinfo{volume}{10} (\bibinfo{year}{2022}), \bibinfo{pages}{1473--1490}.
\newblock


\bibitem[Feng et~al\mbox{.}(2020)]%
        {feng2020deepgini}
\bibfield{author}{\bibinfo{person}{Yang Feng}, \bibinfo{person}{Qingkai Shi}, \bibinfo{person}{Xinyu Gao}, \bibinfo{person}{Jun Wan}, \bibinfo{person}{Chunrong Fang}, {and} \bibinfo{person}{Zhenyu Chen}.} \bibinfo{year}{2020}\natexlab{}.
\newblock \showarticletitle{DeepGini: prioritizing massive tests to enhance the robustness of deep neural networks}. In \bibinfo{booktitle}{\emph{Proceedings of the 29th ACM SIGSOFT International Symposium on Software Testing and Analysis}}. \bibinfo{pages}{177--188}.
\newblock


\bibitem[Gao et~al\mbox{.}(2022)]%
        {gao2022adaptive}
\bibfield{author}{\bibinfo{person}{Xinyu Gao}, \bibinfo{person}{Yang Feng}, \bibinfo{person}{Yining Yin}, \bibinfo{person}{Zixi Liu}, \bibinfo{person}{Zhenyu Chen}, {and} \bibinfo{person}{Baowen Xu}.} \bibinfo{year}{2022}\natexlab{}.
\newblock \showarticletitle{Adaptive test selection for deep neural networks}. In \bibinfo{booktitle}{\emph{Proceedings of the 44th International Conference on Software Engineering}}. \bibinfo{pages}{73--85}.
\newblock


\bibitem[Gao et~al\mbox{.}(2024)]%
        {gao2024multitest}
\bibfield{author}{\bibinfo{person}{Xinyu Gao}, \bibinfo{person}{Zhijie Wang}, \bibinfo{person}{Yang Feng}, \bibinfo{person}{Lei Ma}, \bibinfo{person}{Zhenyu Chen}, {and} \bibinfo{person}{Baowen Xu}.} \bibinfo{year}{2024}\natexlab{}.
\newblock \showarticletitle{MultiTest: Physical-Aware Object Insertion for Testing Multi-sensor Fusion Perception Systems}. In \bibinfo{booktitle}{\emph{Proceedings of the IEEE/ACM 46th International Conference on Software Engineering}}. \bibinfo{pages}{1--13}.
\newblock


\bibitem[Goel and Gupta(2020)]%
        {goel2020robotics}
\bibfield{author}{\bibinfo{person}{Ruchi Goel} {and} \bibinfo{person}{Pooja Gupta}.} \bibinfo{year}{2020}\natexlab{}.
\newblock \bibinfo{booktitle}{\emph{Robotics and Industry 4.0}}.
\newblock \bibinfo{publisher}{Springer International Publishing}, \bibinfo{address}{Cham}, \bibinfo{pages}{157--169}.
\newblock


\bibitem[Gu et~al\mbox{.}(2023)]%
        {gu2023maniskill2}
\bibfield{author}{\bibinfo{person}{Jiayuan Gu}, \bibinfo{person}{Fanbo Xiang}, \bibinfo{person}{Xuanlin Li}, {et~al\mbox{.}}} \bibinfo{year}{2023}\natexlab{}.
\newblock \showarticletitle{ManiSkill2: A Unified Benchmark for Generalizable Manipulation Skills}.
\newblock \bibinfo{journal}{\emph{arXiv preprint arXiv:2302.04659}} (\bibinfo{year}{2023}).
\newblock


\bibitem[Guo et~al\mbox{.}(2023)]%
        {guo2023recent}
\bibfield{author}{\bibinfo{person}{Huihui Guo}, \bibinfo{person}{Fan Wu}, \bibinfo{person}{Yunchuan Qin}, \bibinfo{person}{Ruihui Li}, \bibinfo{person}{Keqin Li}, {and} \bibinfo{person}{Kenli Li}.} \bibinfo{year}{2023}\natexlab{}.
\newblock \showarticletitle{Recent Trends in Task and Motion Planning for Robotics: A Survey}.
\newblock \bibinfo{journal}{\emph{Comput. Surveys}} \bibinfo{volume}{55}, \bibinfo{number}{13s} (\bibinfo{year}{2023}), \bibinfo{pages}{1--36}.
\newblock


\bibitem[Guo et~al\mbox{.}(2020)]%
        {guo2020cps}
\bibfield{author}{\bibinfo{person}{Zhengang Guo}, \bibinfo{person}{Yingfeng Zhang}, \bibinfo{person}{Xibin Zhao}, {and} \bibinfo{person}{Xiaoyu Song}.} \bibinfo{year}{2020}\natexlab{}.
\newblock \showarticletitle{CPS-Based Self-Adaptive Collaborative Control for Smart Production-Logistics Systems}.
\newblock \bibinfo{journal}{\emph{IEEE transactions on cybernetics}} \bibinfo{volume}{51}, \bibinfo{number}{1} (\bibinfo{year}{2020}), \bibinfo{pages}{188--198}.
\newblock


\bibitem[Hamid et~al\mbox{.}(2023)]%
        {hamid2023genaipabench}
\bibfield{author}{\bibinfo{person}{Aamir Hamid}, \bibinfo{person}{Hemanth~Reddy Samidi}, \bibinfo{person}{Tim Finin}, \bibinfo{person}{Primal Pappachan}, {and} \bibinfo{person}{Roberto Yus}.} \bibinfo{year}{2023}\natexlab{}.
\newblock \showarticletitle{GenAIPABench: A Benchmark for Generative AI-based Privacy Assistants}.
\newblock \bibinfo{journal}{\emph{arXiv preprint arXiv:2309.05138}} (\bibinfo{year}{2023}).
\newblock


\bibitem[Han et~al\mbox{.}(2023)]%
        {han2023survey}
\bibfield{author}{\bibinfo{person}{Dong Han}, \bibinfo{person}{Beni Mulyana}, \bibinfo{person}{Vladimir Stankovic}, {and} \bibinfo{person}{Samuel Cheng}.} \bibinfo{year}{2023}\natexlab{}.
\newblock \showarticletitle{A Survey on Deep Reinforcement Learning Algorithms for Robotic Manipulation}.
\newblock \bibinfo{journal}{\emph{Sensors}} \bibinfo{volume}{23}, \bibinfo{number}{7} (\bibinfo{year}{2023}), \bibinfo{pages}{3762}.
\newblock


\bibitem[He et~al\mbox{.}(2020)]%
        {he2020structure}
\bibfield{author}{\bibinfo{person}{Pinjia He}, \bibinfo{person}{Clara Meister}, {and} \bibinfo{person}{Zhendong Su}.} \bibinfo{year}{2020}\natexlab{}.
\newblock \showarticletitle{Structure-Invariant Testing for Machine Translation}. In \bibinfo{booktitle}{\emph{Proceedings of the ACM/IEEE 42nd International Conference on Software Engineering}}. \bibinfo{pages}{961--973}.
\newblock


\bibitem[H{\"o}fer et~al\mbox{.}(2021)]%
        {hofer2021sim2real}
\bibfield{author}{\bibinfo{person}{Sebastian H{\"o}fer}, \bibinfo{person}{Kostas Bekris}, \bibinfo{person}{Ankur Handa}, {et~al\mbox{.}}} \bibinfo{year}{2021}\natexlab{}.
\newblock \showarticletitle{Sim2Real in Robotics and Automation: Applications and Challenges}.
\newblock \bibinfo{journal}{\emph{IEEE Transactions on Automation Science and Engineering}} \bibinfo{volume}{18}, \bibinfo{number}{2} (\bibinfo{year}{2021}), \bibinfo{pages}{398--400}.
\newblock


\bibitem[Holland et~al\mbox{.}(2021)]%
        {holland2021service}
\bibfield{author}{\bibinfo{person}{Jane Holland}, \bibinfo{person}{Liz Kingston}, \bibinfo{person}{Conor McCarthy}, \bibinfo{person}{Eddie Armstrong}, \bibinfo{person}{Peter O’Dwyer}, \bibinfo{person}{Fionn Merz}, {and} \bibinfo{person}{Mark McConnell}.} \bibinfo{year}{2021}\natexlab{}.
\newblock \showarticletitle{Service Robots in the Healthcare Sector}.
\newblock \bibinfo{journal}{\emph{Robotics}} \bibinfo{volume}{10}, \bibinfo{number}{1} (\bibinfo{year}{2021}), \bibinfo{pages}{47}.
\newblock


\bibitem[Hu et~al\mbox{.}(2023)]%
        {hu2023aries}
\bibfield{author}{\bibinfo{person}{Qiang Hu}, \bibinfo{person}{Yuejun Guo}, \bibinfo{person}{Xiaofei Xie}, \bibinfo{person}{Maxime Cordy}, \bibinfo{person}{Mike Papadakis}, \bibinfo{person}{Lei Ma}, {and} \bibinfo{person}{Yves Le~Traon}.} \bibinfo{year}{2023}\natexlab{}.
\newblock \showarticletitle{Aries: Efficient Testing of Deep Neural Networks via Labeling-Free Accuracy Estimation}. In \bibinfo{booktitle}{\emph{2023 IEEE/ACM 45th International Conference on Software Engineering (ICSE)}}. IEEE, \bibinfo{pages}{1776--1787}.
\newblock


\bibitem[Hu et~al\mbox{.}(2021)]%
        {hu2021coverage}
\bibfield{author}{\bibinfo{person}{Zhisheng Hu}, \bibinfo{person}{Shengjian Guo}, \bibinfo{person}{Zhenyu Zhong}, {and} \bibinfo{person}{Kang Li}.} \bibinfo{year}{2021}\natexlab{}.
\newblock \showarticletitle{Coverage-based Scene Fuzzing for Virtual Autonomous Driving Testing}.
\newblock \bibinfo{journal}{\emph{arXiv preprint arXiv:2106.00873}} (\bibinfo{year}{2021}).
\newblock


\bibitem[Huang et~al\mbox{.}(2023)]%
        {huang2023instruct2act}
\bibfield{author}{\bibinfo{person}{Siyuan Huang}, \bibinfo{person}{Zhengkai Jiang}, \bibinfo{person}{Hao Dong}, \bibinfo{person}{Yu Qiao}, \bibinfo{person}{Peng Gao}, {and} \bibinfo{person}{Hongsheng Li}.} \bibinfo{year}{2023}\natexlab{}.
\newblock \showarticletitle{Instruct2Act: Mapping Multi-modality Instructions to Robotic Actions with Large Language Model}.
\newblock \bibinfo{journal}{\emph{arXiv preprint arXiv:2305.11176}} (\bibinfo{year}{2023}).
\newblock


\bibitem[Huang et~al\mbox{.}(2024)]%
        {huang2024active}
\bibfield{author}{\bibinfo{person}{Yuheng Huang}, \bibinfo{person}{Jiayang Song}, \bibinfo{person}{Qiang Hu}, \bibinfo{person}{Felix Juefei-Xu}, {and} \bibinfo{person}{Lei Ma}.} \bibinfo{year}{2024}\natexlab{}.
\newblock \showarticletitle{Active Testing of Large Language Model via Multi-Stage Sampling}.
\newblock \bibinfo{journal}{\emph{arXiv preprint arXiv:2408.03573}} (\bibinfo{year}{2024}).
\newblock


\bibitem[Huang et~al\mbox{.}(2025)]%
        {huang2023look}
\bibfield{author}{\bibinfo{person}{Yuheng Huang}, \bibinfo{person}{Jiayang Song}, \bibinfo{person}{Zhijie Wang}, {et~al\mbox{.}}} \bibinfo{year}{2025}\natexlab{}.
\newblock \showarticletitle{Look Before You Leap: An Exploratory Study of Uncertainty Analysis for Large Language Models}.
\newblock \bibinfo{journal}{\emph{IEEE Transactions on Software Engineering}} \bibinfo{volume}{51}, \bibinfo{number}{2} (\bibinfo{year}{2025}), \bibinfo{pages}{413--429}.
\newblock


\bibitem[Jimenez et~al\mbox{.}(2024)]%
        {jimenez2023swe}
\bibfield{author}{\bibinfo{person}{Carlos~E Jimenez}, \bibinfo{person}{John Yang}, \bibinfo{person}{Alexander Wettig}, \bibinfo{person}{Shunyu Yao}, \bibinfo{person}{Kexin Pei}, \bibinfo{person}{Ofir Press}, {and} \bibinfo{person}{Karthik Narasimhan}.} \bibinfo{year}{2024}\natexlab{}.
\newblock \showarticletitle{SWE-bench: Can Language Models Resolve Real-World GitHub Issues?}. In \bibinfo{booktitle}{\emph{The Twelfth International Conference on Learning Representations}}.
\newblock


\bibitem[Kim et~al\mbox{.}(2024)]%
        {kim2024openvla}
\bibfield{author}{\bibinfo{person}{Moo~Jin Kim}, \bibinfo{person}{Karl Pertsch}, \bibinfo{person}{Siddharth Karamcheti}, {et~al\mbox{.}}} \bibinfo{year}{2024}\natexlab{}.
\newblock \showarticletitle{OpenVLA: An Open-Source Vision-Language-Action Model}.
\newblock \bibinfo{journal}{\emph{arXiv preprint arXiv:2406.09246}} (\bibinfo{year}{2024}).
\newblock


\bibitem[Kyrarini et~al\mbox{.}(2021)]%
        {kyrarini2021survey}
\bibfield{author}{\bibinfo{person}{Maria Kyrarini}, \bibinfo{person}{Fotios Lygerakis}, \bibinfo{person}{Akilesh Rajavenkatanarayanan}, \bibinfo{person}{Christos Sevastopoulos}, \bibinfo{person}{Harish~Ram Nambiappan}, \bibinfo{person}{Kodur~Krishna Chaitanya}, \bibinfo{person}{Ashwin~Ramesh Babu}, \bibinfo{person}{Joanne Mathew}, {and} \bibinfo{person}{Fillia Makedon}.} \bibinfo{year}{2021}\natexlab{}.
\newblock \showarticletitle{A Survey of Robots in Healthcare}.
\newblock \bibinfo{journal}{\emph{Technologies}} \bibinfo{volume}{9}, \bibinfo{number}{1} (\bibinfo{year}{2021}), \bibinfo{pages}{8}.
\newblock


\bibitem[Lee et~al\mbox{.}(2024)]%
        {lee2024automated}
\bibfield{author}{\bibinfo{person}{Jaeseong Lee}, \bibinfo{person}{Simin Chen}, \bibinfo{person}{Austin Mordahl}, \bibinfo{person}{Cong Liu}, \bibinfo{person}{Wei Yang}, {and} \bibinfo{person}{Shiyi Wei}.} \bibinfo{year}{2024}\natexlab{}.
\newblock \showarticletitle{Automated Testing Linguistic Capabilities of NLP Models}.
\newblock \bibinfo{journal}{\emph{ACM Transactions on Software Engineering and Methodology}} (\bibinfo{year}{2024}).
\newblock


\bibitem[Lee et~al\mbox{.}(2023)]%
        {lee2023fuzzing}
\bibfield{author}{\bibinfo{person}{Jaekwon Lee}, \bibinfo{person}{Enrico Vigan{\`o}}, \bibinfo{person}{Oscar Cornejo}, \bibinfo{person}{Fabrizio Pastore}, {and} \bibinfo{person}{Lionel Briand}.} \bibinfo{year}{2023}\natexlab{}.
\newblock \showarticletitle{Fuzzing for CPS Mutation Testing}. In \bibinfo{booktitle}{\emph{2023 38th IEEE/ACM International Conference on Automated Software Engineering (ASE)}}. IEEE, \bibinfo{pages}{1377--1389}.
\newblock


\bibitem[Li et~al\mbox{.}(2024b)]%
        {li2024evaluating}
\bibfield{author}{\bibinfo{person}{Xuanlin Li}, \bibinfo{person}{Kyle Hsu}, \bibinfo{person}{Jiayuan Gu}, {et~al\mbox{.}}} \bibinfo{year}{2024}\natexlab{b}.
\newblock \showarticletitle{Evaluating Real-World Robot Manipulation Policies in Simulation}. In \bibinfo{booktitle}{\emph{8th Annual Conference on Robot Learning}}.
\newblock


\bibitem[Li et~al\mbox{.}(2024c)]%
        {li2023vision}
\bibfield{author}{\bibinfo{person}{Xinghang Li}, \bibinfo{person}{Minghuan Liu}, \bibinfo{person}{Hanbo Zhang}, {et~al\mbox{.}}} \bibinfo{year}{2024}\natexlab{c}.
\newblock \showarticletitle{Vision-Language Foundation Models as Effective Robot Imitators}. In \bibinfo{booktitle}{\emph{The Twelfth International Conference on Learning Representations}}.
\newblock


\bibitem[Li et~al\mbox{.}(2025)]%
        {li2024llara}
\bibfield{author}{\bibinfo{person}{Xiang Li}, \bibinfo{person}{Cristina Mata}, \bibinfo{person}{Jongwoo Park}, {et~al\mbox{.}}} \bibinfo{year}{2025}\natexlab{}.
\newblock \showarticletitle{LLaRA: Supercharging Robot Learning Data for Vision-Language Policy}. In \bibinfo{booktitle}{\emph{The Thirteenth International Conference on Learning Representations}}.
\newblock


\bibitem[Li et~al\mbox{.}(2024a)]%
        {li2024prioritizing}
\bibfield{author}{\bibinfo{person}{Yinghua Li}, \bibinfo{person}{Xueqi Dang}, \bibinfo{person}{Lei Ma}, \bibinfo{person}{Jacques Klein}, {and} \bibinfo{person}{Tegawend{\'e}~F Bissyand{\'e}}.} \bibinfo{year}{2024}\natexlab{a}.
\newblock \showarticletitle{Prioritizing test cases for deep learning-based video classifiers}.
\newblock \bibinfo{journal}{\emph{Empirical Software Engineering}} \bibinfo{volume}{29}, \bibinfo{number}{5} (\bibinfo{year}{2024}), \bibinfo{pages}{111}.
\newblock


\bibitem[Liang et~al\mbox{.}(2023b)]%
        {liang2023code}
\bibfield{author}{\bibinfo{person}{Jacky Liang}, \bibinfo{person}{Wenlong Huang}, {et~al\mbox{.}}} \bibinfo{year}{2023}\natexlab{b}.
\newblock \showarticletitle{Code as Policies: Language Model Programs for Embodied Control}. In \bibinfo{booktitle}{\emph{2023 IEEE International Conference on Robotics and Automation (ICRA)}}. IEEE, \bibinfo{pages}{9493--9500}.
\newblock


\bibitem[Liang et~al\mbox{.}(2023a)]%
        {liang2023holistic}
\bibfield{author}{\bibinfo{person}{Percy Liang}, \bibinfo{person}{Rishi Bommasani}, \bibinfo{person}{Tony Lee}, {et~al\mbox{.}}} \bibinfo{year}{2023}\natexlab{a}.
\newblock \showarticletitle{Holistic Evaluation of Language Models}.
\newblock \bibinfo{journal}{\emph{Transactions on Machine Learning Research}} (\bibinfo{year}{2023}).
\newblock
\showISSN{2835-8856}


\bibitem[Lin et~al\mbox{.}(2022)]%
        {lin2021truthfulqa}
\bibfield{author}{\bibinfo{person}{Stephanie Lin}, \bibinfo{person}{Jacob Hilton}, {and} \bibinfo{person}{Owain Evans}.} \bibinfo{year}{2022}\natexlab{}.
\newblock \showarticletitle{{T}ruthful{QA}: Measuring How Models Mimic Human Falsehoods}. In \bibinfo{booktitle}{\emph{Proceedings of the 60th Annual Meeting of the Association for Computational Linguistics}}. \bibinfo{pages}{3214--3252}.
\newblock


\bibitem[Liu et~al\mbox{.}(2024a)]%
        {liu2024visual}
\bibfield{author}{\bibinfo{person}{Haotian Liu}, \bibinfo{person}{Chunyuan Li}, \bibinfo{person}{Qingyang Wu}, {and} \bibinfo{person}{Yong~Jae Lee}.} \bibinfo{year}{2024}\natexlab{a}.
\newblock \showarticletitle{Visual Instruction Tuning}.
\newblock \bibinfo{journal}{\emph{Advances in neural information processing systems}}  \bibinfo{volume}{36} (\bibinfo{year}{2024}).
\newblock


\bibitem[Liu et~al\mbox{.}(2024b)]%
        {liu2024your}
\bibfield{author}{\bibinfo{person}{Jiawei Liu}, \bibinfo{person}{Chunqiu~Steven Xia}, \bibinfo{person}{Yuyao Wang}, {and} \bibinfo{person}{Lingming Zhang}.} \bibinfo{year}{2024}\natexlab{b}.
\newblock \showarticletitle{Is Your Code Generated by ChatGPT Really Correct? Rigorous Evaluation of Large Language Models for Code Generation}.
\newblock \bibinfo{journal}{\emph{Advances in Neural Information Processing Systems}}  \bibinfo{volume}{36} (\bibinfo{year}{2024}).
\newblock


\bibitem[Menghi et~al\mbox{.}(2020)]%
        {menghi2020approximation}
\bibfield{author}{\bibinfo{person}{Claudio Menghi}, \bibinfo{person}{Shiva Nejati}, \bibinfo{person}{Lionel Briand}, {and} \bibinfo{person}{Yago~Isasi Parache}.} \bibinfo{year}{2020}\natexlab{}.
\newblock \showarticletitle{Approximation-Refinement Testing of Compute-Intensive Cyber-Physical Models: An Approach Based on System Identification}. In \bibinfo{booktitle}{\emph{Proceedings of the ACM/IEEE 42nd International Conference on Software Engineering}}. \bibinfo{pages}{372--384}.
\newblock


\bibitem[Mukherjee and Awadallah(2020)]%
        {mukherjee2020uncertainty}
\bibfield{author}{\bibinfo{person}{Subhabrata Mukherjee} {and} \bibinfo{person}{Ahmed Awadallah}.} \bibinfo{year}{2020}\natexlab{}.
\newblock \showarticletitle{Uncertainty-aware Self-training for Few-shot Text Classification}. In \bibinfo{booktitle}{\emph{Advances in Neural Information Processing Systems}}, Vol.~\bibinfo{volume}{33}. \bibinfo{publisher}{Curran Associates, Inc.}, \bibinfo{pages}{21199--21212}.
\newblock


\bibitem[Murray et~al\mbox{.}(2017)]%
        {murray2017mathematical}
\bibfield{author}{\bibinfo{person}{Richard~M Murray}, \bibinfo{person}{Zexiang Li}, {and} \bibinfo{person}{S~Shankar Sastry}.} \bibinfo{year}{2017}\natexlab{}.
\newblock \bibinfo{booktitle}{\emph{A Mathematical Introduction to Robotic Manipulation}}.
\newblock \bibinfo{publisher}{CRC press}.
\newblock


\bibitem[Nair et~al\mbox{.}(2023)]%
        {nair2022r3m}
\bibfield{author}{\bibinfo{person}{Suraj Nair}, \bibinfo{person}{Aravind Rajeswaran}, \bibinfo{person}{Vikash Kumar}, \bibinfo{person}{Chelsea Finn}, {and} \bibinfo{person}{Abhinav Gupta}.} \bibinfo{year}{2023}\natexlab{}.
\newblock \showarticletitle{R3M: A Universal Visual Representation for Robot Manipulation}. In \bibinfo{booktitle}{\emph{Proceedings of The 6th Conference on Robot Learning}} \emph{(\bibinfo{series}{Proceedings of Machine Learning Research}, Vol.~\bibinfo{volume}{205})}. \bibinfo{publisher}{PMLR}, \bibinfo{pages}{892--909}.
\newblock


\bibitem[Oquab et~al\mbox{.}(2023)]%
        {oquabdinov2}
\bibfield{author}{\bibinfo{person}{Maxime Oquab}, \bibinfo{person}{Timoth{\'e}e Darcet}, \bibinfo{person}{Th{\'e}o Moutakanni}, {et~al\mbox{.}}} \bibinfo{year}{2023}\natexlab{}.
\newblock \showarticletitle{DINOv2: Learning Robust Visual Features without Supervision}.
\newblock \bibinfo{journal}{\emph{Transactions on Machine Learning Research}} (\bibinfo{year}{2023}).
\newblock


\bibitem[Orr and Dutta(2023)]%
        {orr2023multi}
\bibfield{author}{\bibinfo{person}{James Orr} {and} \bibinfo{person}{Ayan Dutta}.} \bibinfo{year}{2023}\natexlab{}.
\newblock \showarticletitle{Multi-Agent Deep Reinforcement Learning for Multi-Robot Applications: A Survey}.
\newblock \bibinfo{journal}{\emph{Sensors}} \bibinfo{volume}{23}, \bibinfo{number}{7} (\bibinfo{year}{2023}), \bibinfo{pages}{3625}.
\newblock


\bibitem[Padalkar et~al\mbox{.}(2023)]%
        {padalkar2023open}
\bibfield{author}{\bibinfo{person}{Abhishek Padalkar}, \bibinfo{person}{Acorn Pooley}, \bibinfo{person}{Ajinkya Jain}, {et~al\mbox{.}}} \bibinfo{year}{2023}\natexlab{}.
\newblock \showarticletitle{Open X-Embodiment: Robotic Learning Datasets and RT-X Models}.
\newblock \bibinfo{journal}{\emph{arXiv preprint arXiv:2310.08864}} (\bibinfo{year}{2023}).
\newblock


\bibitem[Pan et~al\mbox{.}(2024)]%
        {pan2024lost}
\bibfield{author}{\bibinfo{person}{Rangeet Pan}, \bibinfo{person}{Ali~Reza Ibrahimzada}, {et~al\mbox{.}}} \bibinfo{year}{2024}\natexlab{}.
\newblock \showarticletitle{Lost in Translation: A Study of Bugs Introduced by Large Language Models while Translating Code}. In \bibinfo{booktitle}{\emph{Proceedings of the IEEE/ACM 46th International Conference on Software Engineering}}.
\newblock


\bibitem[Perez et~al\mbox{.}(2018)]%
        {perez2018film}
\bibfield{author}{\bibinfo{person}{Ethan Perez}, \bibinfo{person}{Florian Strub}, \bibinfo{person}{Harm De~Vries}, \bibinfo{person}{Vincent Dumoulin}, {and} \bibinfo{person}{Aaron Courville}.} \bibinfo{year}{2018}\natexlab{}.
\newblock \showarticletitle{FiLM: Visual Reasoning with a General Conditioning Layer}. In \bibinfo{booktitle}{\emph{Proceedings of the AAAI conference on artificial intelligence}}, Vol.~\bibinfo{volume}{32}.
\newblock


\bibitem[Rayhan(2023)]%
        {rayhan2023artificial}
\bibfield{author}{\bibinfo{person}{Abu Rayhan}.} \bibinfo{year}{2023}\natexlab{}.
\newblock \bibinfo{title}{Artificial intelligence in robotics: From automation to autonomous systems}.
\newblock


\bibitem[Russakovsky et~al\mbox{.}(2015)]%
        {russakovsky2015imagenet}
\bibfield{author}{\bibinfo{person}{Olga Russakovsky}, \bibinfo{person}{Jia Deng}, \bibinfo{person}{Hao Su}, {et~al\mbox{.}}} \bibinfo{year}{2015}\natexlab{}.
\newblock \showarticletitle{ImageNet Large Scale Visual Recognition Challenge}.
\newblock \bibinfo{journal}{\emph{International journal of computer vision}}  \bibinfo{volume}{115} (\bibinfo{year}{2015}), \bibinfo{pages}{211--252}.
\newblock


\bibitem[Shamout et~al\mbox{.}(2022)]%
        {shamout2022conceptual}
\bibfield{author}{\bibinfo{person}{Mohamed Shamout}, \bibinfo{person}{Rabeb Ben-Abdallah}, {et~al\mbox{.}}} \bibinfo{year}{2022}\natexlab{}.
\newblock \showarticletitle{A Conceptual Model for the Adoption of Autonomous Robots in the Supply Chain and Logistics Industry}.
\newblock \bibinfo{journal}{\emph{Uncertain Supply Chain Management}} \bibinfo{volume}{10}, \bibinfo{number}{2} (\bibinfo{year}{2022}), \bibinfo{pages}{577--592}.
\newblock


\bibitem[Siciliano(2008)]%
        {siciliano2008springer}
\bibfield{author}{\bibinfo{person}{B Siciliano}.} \bibinfo{year}{2008}\natexlab{}.
\newblock \bibinfo{booktitle}{\emph{Springer Handbook of Robotics}}.
\newblock \bibinfo{publisher}{Springer}.
\newblock


\bibitem[Singh et~al\mbox{.}(2023)]%
        {singh2023progprompt}
\bibfield{author}{\bibinfo{person}{Ishika Singh}, \bibinfo{person}{Valts Blukis}, \bibinfo{person}{Arsalan Mousavian}, {et~al\mbox{.}}} \bibinfo{year}{2023}\natexlab{}.
\newblock \showarticletitle{ProgPrompt: Generating Situated Robot Task Plans using Large Language Models}. In \bibinfo{booktitle}{\emph{2023 IEEE International Conference on Robotics and Automation (ICRA)}}. IEEE, \bibinfo{pages}{11523--11530}.
\newblock


\bibitem[Song et~al\mbox{.}(2023a)]%
        {song2023mathtt}
\bibfield{author}{\bibinfo{person}{Jiayang Song}, \bibinfo{person}{Xuan Xie}, {and} \bibinfo{person}{Lei Ma}.} \bibinfo{year}{2023}\natexlab{a}.
\newblock \showarticletitle{SIEGE: A Semantics-Guided Safety Enhancement Framework for AI-Enabled Cyber-Physical Systems}.
\newblock \bibinfo{journal}{\emph{IEEE Transactions on Software Engineering}} \bibinfo{volume}{49}, \bibinfo{number}{8} (\bibinfo{year}{2023}), \bibinfo{pages}{4058--4080}.
\newblock


\bibitem[Song et~al\mbox{.}(2023b)]%
        {song2023self}
\bibfield{author}{\bibinfo{person}{Jiayang Song}, \bibinfo{person}{Zhehua Zhou}, \bibinfo{person}{Jiawei Liu}, \bibinfo{person}{Chunrong Fang}, \bibinfo{person}{Zhan Shu}, {and} \bibinfo{person}{Lei Ma}.} \bibinfo{year}{2023}\natexlab{b}.
\newblock \showarticletitle{Self-Refined Large Language Model as Automated Reward Function Designer for Deep Reinforcement Learning in Robotics}.
\newblock \bibinfo{journal}{\emph{arXiv preprint arXiv:2309.06687}} (\bibinfo{year}{2023}).
\newblock


\bibitem[Soori et~al\mbox{.}(2023)]%
        {soori2023artificial}
\bibfield{author}{\bibinfo{person}{Mohsen Soori}, \bibinfo{person}{Behrooz Arezoo}, {and} \bibinfo{person}{Roza Dastres}.} \bibinfo{year}{2023}\natexlab{}.
\newblock \showarticletitle{Artificial intelligence, machine learning and deep learning in advanced robotics, a review}.
\newblock \bibinfo{journal}{\emph{Cognitive Robotics}}  \bibinfo{volume}{3} (\bibinfo{year}{2023}), \bibinfo{pages}{54--70}.
\newblock


\bibitem[Stocco and Tonella(2022)]%
        {stocco2022confidence}
\bibfield{author}{\bibinfo{person}{Andrea Stocco} {and} \bibinfo{person}{Paolo Tonella}.} \bibinfo{year}{2022}\natexlab{}.
\newblock \showarticletitle{Confidence-driven weighted retraining for predicting safety-critical failures in autonomous driving systems}.
\newblock \bibinfo{journal}{\emph{Journal of Software: Evolution and Process}} \bibinfo{volume}{34}, \bibinfo{number}{10} (\bibinfo{year}{2022}), \bibinfo{pages}{e2386}.
\newblock


\bibitem[Stone et~al\mbox{.}(2023)]%
        {stone2023open}
\bibfield{author}{\bibinfo{person}{Austin Stone} {et~al\mbox{.}}} \bibinfo{year}{2023}\natexlab{}.
\newblock \showarticletitle{Open-World Object Manipulation using Pre-trained Vision-Language Models}. In \bibinfo{booktitle}{\emph{Proceedings of The 7th Conference on Robot Learning}} \emph{(\bibinfo{series}{Proceedings of Machine Learning Research}, Vol.~\bibinfo{volume}{229})}. \bibinfo{publisher}{PMLR}, \bibinfo{pages}{3397--3417}.
\newblock


\bibitem[Sun et~al\mbox{.}(2024)]%
        {sun2024trustllm}
\bibfield{author}{\bibinfo{person}{Lichao Sun}, \bibinfo{person}{Yue Huang}, {et~al\mbox{.}}} \bibinfo{year}{2024}\natexlab{}.
\newblock \showarticletitle{TrustLLM: Trustworthiness in Large Language Models}. In \bibinfo{booktitle}{\emph{Proceedings of the 41st International Conference on Machine Learning}} \emph{(\bibinfo{series}{Proceedings of Machine Learning Research}, Vol.~\bibinfo{volume}{235})}. \bibinfo{publisher}{PMLR}, \bibinfo{pages}{20166--20270}.
\newblock


\bibitem[Sun et~al\mbox{.}(2020)]%
        {sun2020automatic}
\bibfield{author}{\bibinfo{person}{Zeyu Sun}, \bibinfo{person}{Jie~M Zhang}, \bibinfo{person}{Mark Harman}, \bibinfo{person}{Mike Papadakis}, {and} \bibinfo{person}{Lu Zhang}.} \bibinfo{year}{2020}\natexlab{}.
\newblock \showarticletitle{Automatic Testing and Improvement of Machine Translation}. In \bibinfo{booktitle}{\emph{Proceedings of the ACM/IEEE 42nd International Conference on Software Engineering}}. \bibinfo{pages}{974--985}.
\newblock


\bibitem[Tan and Le(2019)]%
        {tan2019efficientnet}
\bibfield{author}{\bibinfo{person}{Mingxing Tan} {and} \bibinfo{person}{Quoc Le}.} \bibinfo{year}{2019}\natexlab{}.
\newblock \showarticletitle{{E}fficient{N}et: Rethinking Model Scaling for Convolutional Neural Networks}. In \bibinfo{booktitle}{\emph{Proceedings of the 36th International Conference on Machine Learning}} \emph{(\bibinfo{series}{Proceedings of Machine Learning Research}, Vol.~\bibinfo{volume}{97})}, \bibfield{editor}{\bibinfo{person}{Kamalika Chaudhuri} {and} \bibinfo{person}{Ruslan Salakhutdinov}} (Eds.). \bibinfo{publisher}{PMLR}, \bibinfo{pages}{6105--6114}.
\newblock


\bibitem[Team et~al\mbox{.}(2024)]%
        {team2024octo}
\bibfield{author}{\bibinfo{person}{Octo~Model Team}, \bibinfo{person}{Dibya Ghosh}, \bibinfo{person}{Homer Walke}, {et~al\mbox{.}}} \bibinfo{year}{2024}\natexlab{}.
\newblock \showarticletitle{Octo: An Open-Source Generalist Robot Policy}.
\newblock \bibinfo{journal}{\emph{arXiv preprint arXiv:2405.12213}} (\bibinfo{year}{2024}).
\newblock


\bibitem[Touvron et~al\mbox{.}(2023)]%
        {touvron2023llama}
\bibfield{author}{\bibinfo{person}{Hugo Touvron}, \bibinfo{person}{Louis Martin}, \bibinfo{person}{Kevin Stone}, {et~al\mbox{.}}} \bibinfo{year}{2023}\natexlab{}.
\newblock \showarticletitle{Llama 2: Open Foundation and Fine-Tuned Chat Models}.
\newblock \bibinfo{journal}{\emph{arXiv preprint arXiv:2307.09288}} (\bibinfo{year}{2023}).
\newblock


\bibitem[Wakchaure et~al\mbox{.}(2023)]%
        {wakchaure2023application}
\bibfield{author}{\bibinfo{person}{Manas Wakchaure}, \bibinfo{person}{BK Patle}, {and} \bibinfo{person}{AK Mahindrakar}.} \bibinfo{year}{2023}\natexlab{}.
\newblock \showarticletitle{Application of AI techniques and robotics in agriculture: A review}.
\newblock \bibinfo{journal}{\emph{Artificial Intelligence in the Life Sciences}}  \bibinfo{volume}{3} (\bibinfo{year}{2023}), \bibinfo{pages}{100057}.
\newblock


\bibitem[Wan et~al\mbox{.}(2023)]%
        {wan2023biasasker}
\bibfield{author}{\bibinfo{person}{Yuxuan Wan}, \bibinfo{person}{Wenxuan Wang}, \bibinfo{person}{Pinjia He}, \bibinfo{person}{Jiazhen Gu}, \bibinfo{person}{Haonan Bai}, {and} \bibinfo{person}{Michael~R Lyu}.} \bibinfo{year}{2023}\natexlab{}.
\newblock \showarticletitle{BiasAsker: Measuring the Bias in Conversational AI System}. In \bibinfo{booktitle}{\emph{Proceedings of the 31st ACM Joint European Software Engineering Conference and Symposium on the Foundations of Software Engineering}}. \bibinfo{pages}{515--527}.
\newblock


\bibitem[Wang et~al\mbox{.}(2023)]%
        {wang2023decodingtrust}
\bibfield{author}{\bibinfo{person}{Boxin Wang}, \bibinfo{person}{Weixin Chen}, \bibinfo{person}{Hengzhi Pei}, {et~al\mbox{.}}} \bibinfo{year}{2023}\natexlab{}.
\newblock \showarticletitle{DecodingTrust: A Comprehensive Assessment of Trustworthiness in {GPT} Models}. In \bibinfo{booktitle}{\emph{Thirty-seventh Conference on Neural Information Processing Systems Datasets and Benchmarks Track}}.
\newblock


\bibitem[Wang et~al\mbox{.}(2024)]%
        {wang2024mortar}
\bibfield{author}{\bibinfo{person}{Renzhi Wang}, \bibinfo{person}{Zhehua Zhou}, \bibinfo{person}{Jiayang Song}, \bibinfo{person}{Xuan Xie}, \bibinfo{person}{Xiaofei Xie}, {and} \bibinfo{person}{Lei Ma}.} \bibinfo{year}{2024}\natexlab{}.
\newblock \showarticletitle{MORTAR: A Model-based Runtime Action Repair Framework for AI-enabled Cyber-Physical Systems}.
\newblock \bibinfo{journal}{\emph{arXiv preprint arXiv:2408.03892}} (\bibinfo{year}{2024}).
\newblock


\bibitem[Wang and Su(2020)]%
        {wang2020metamorphic}
\bibfield{author}{\bibinfo{person}{Shuai Wang} {and} \bibinfo{person}{Zhendong Su}.} \bibinfo{year}{2020}\natexlab{}.
\newblock \showarticletitle{Metamorphic Object Insertion for Testing Object Detection Systems}. In \bibinfo{booktitle}{\emph{Proceedings of the 35th IEEE/ACM International Conference on Automated Software Engineering}}. \bibinfo{pages}{1053--1065}.
\newblock


\bibitem[Wang et~al\mbox{.}(2025)]%
        {wang2025towards}
\bibfield{author}{\bibinfo{person}{Zhijie Wang}, \bibinfo{person}{Zijie Zhou}, \bibinfo{person}{Da Song}, \bibinfo{person}{Yuheng Huang}, \bibinfo{person}{Shengmai Chen}, \bibinfo{person}{Lei Ma}, {and} \bibinfo{person}{Tianyi Zhang}.} \bibinfo{year}{2025}\natexlab{}.
\newblock \showarticletitle{Towards Understanding the Characteristics of Code Generation Errors Made by Large Language Models}. In \bibinfo{booktitle}{\emph{Proceedings of the IEEE/ACM 47th International Conference on software Engineering (ICSE '25)}}.
\newblock


\bibitem[Wei et~al\mbox{.}(2022)]%
        {wei2022chain}
\bibfield{author}{\bibinfo{person}{Jason Wei}, \bibinfo{person}{Xuezhi Wang}, \bibinfo{person}{Dale Schuurmans}, {et~al\mbox{.}}} \bibinfo{year}{2022}\natexlab{}.
\newblock \showarticletitle{Chain-of-Thought Prompting Elicits Reasoning in Large Language Models}. In \bibinfo{booktitle}{\emph{Advances in Neural Information Processing Systems}}, Vol.~\bibinfo{volume}{35}. \bibinfo{publisher}{Curran Associates, Inc.}, \bibinfo{pages}{24824--24837}.
\newblock


\bibitem[Xia et~al\mbox{.}(2024)]%
        {xia2024agentless}
\bibfield{author}{\bibinfo{person}{Chunqiu~Steven Xia}, \bibinfo{person}{Yinlin Deng}, \bibinfo{person}{Soren Dunn}, {and} \bibinfo{person}{Lingming Zhang}.} \bibinfo{year}{2024}\natexlab{}.
\newblock \showarticletitle{Agentless: Demystifying LLM-based Software Engineering Agents}.
\newblock \bibinfo{journal}{\emph{arXiv preprint arXiv:2407.01489}} (\bibinfo{year}{2024}).
\newblock


\bibitem[Xiao et~al\mbox{.}(2023)]%
        {xiao2023leap}
\bibfield{author}{\bibinfo{person}{Mingxuan Xiao}, \bibinfo{person}{Yan Xiao}, {et~al\mbox{.}}} \bibinfo{year}{2023}\natexlab{}.
\newblock \showarticletitle{LEAP: Efficient and Automated Test Method for NLP Software}. In \bibinfo{booktitle}{\emph{Proceedings of the 38th IEEE/ACM International Conference on Automated Software Engineering}}. \bibinfo{pages}{1136--1148}.
\newblock


\bibitem[Xie et~al\mbox{.}(2024)]%
        {xie2024online}
\bibfield{author}{\bibinfo{person}{Xuan Xie}, \bibinfo{person}{Jiayang Song}, \bibinfo{person}{Zhehua Zhou}, \bibinfo{person}{Yuheng Huang}, \bibinfo{person}{Da Song}, {and} \bibinfo{person}{Lei Ma}.} \bibinfo{year}{2024}\natexlab{}.
\newblock \showarticletitle{Online Safety Analysis for LLMs: a Benchmark, an Assessment, and a Path Forward}.
\newblock \bibinfo{journal}{\emph{arXiv preprint arXiv:2404.08517}} (\bibinfo{year}{2024}).
\newblock


\bibitem[Yang et~al\mbox{.}(2024)]%
        {yang2024swe}
\bibfield{author}{\bibinfo{person}{John Yang}, \bibinfo{person}{Carlos~E. Jimenez}, {et~al\mbox{.}}} \bibinfo{year}{2024}\natexlab{}.
\newblock \showarticletitle{SWE-agent: Agent-Computer Interfaces Enable Automated Software Engineering}. In \bibinfo{booktitle}{\emph{Advances in Neural Information Processing Systems}}, Vol.~\bibinfo{volume}{37}. \bibinfo{publisher}{Curran Associates, Inc.}, \bibinfo{pages}{50528--50652}.
\newblock


\bibitem[Yao et~al\mbox{.}(2023)]%
        {yao2024tree}
\bibfield{author}{\bibinfo{person}{Shunyu Yao}, \bibinfo{person}{Dian Yu}, \bibinfo{person}{Jeffrey Zhao}, {et~al\mbox{.}}} \bibinfo{year}{2023}\natexlab{}.
\newblock \showarticletitle{Tree of Thoughts: Deliberate Problem Solving with Large Language Models}. In \bibinfo{booktitle}{\emph{Advances in Neural Information Processing Systems}}, Vol.~\bibinfo{volume}{36}. \bibinfo{publisher}{Curran Associates, Inc.}, \bibinfo{pages}{11809--11822}.
\newblock


\bibitem[Yohanandhan et~al\mbox{.}(2020)]%
        {yohanandhan2020cyber}
\bibfield{author}{\bibinfo{person}{Rajaa~Vikhram Yohanandhan}, \bibinfo{person}{Rajvikram~Madurai Elavarasan}, {et~al\mbox{.}}} \bibinfo{year}{2020}\natexlab{}.
\newblock \showarticletitle{Cyber-Physical Power System (CPPS): A Review on Modeling, Simulation, and Analysis With Cyber Security Applications}.
\newblock \bibinfo{journal}{\emph{IEEE Access}}  \bibinfo{volume}{8} (\bibinfo{year}{2020}), \bibinfo{pages}{151019--151064}.
\newblock


\bibitem[Yu et~al\mbox{.}(2022)]%
        {yu2022automated}
\bibfield{author}{\bibinfo{person}{Boxi Yu}, \bibinfo{person}{Zhiqing Zhong}, \bibinfo{person}{Xinran Qin}, \bibinfo{person}{Jiayi Yao}, \bibinfo{person}{Yuancheng Wang}, {and} \bibinfo{person}{Pinjia He}.} \bibinfo{year}{2022}\natexlab{}.
\newblock \showarticletitle{Automated testing of image captioning systems}. In \bibinfo{booktitle}{\emph{Proceedings of the 31st ACM SIGSOFT International Symposium on Software Testing and Analysis}}. \bibinfo{pages}{467--479}.
\newblock


\bibitem[Yu et~al\mbox{.}(2024)]%
        {yu2024codereval}
\bibfield{author}{\bibinfo{person}{Hao Yu}, \bibinfo{person}{Bo Shen}, \bibinfo{person}{Dezhi Ran}, \bibinfo{person}{Jiaxin Zhang}, \bibinfo{person}{Qi Zhang}, \bibinfo{person}{Yuchi Ma}, \bibinfo{person}{Guangtai Liang}, \bibinfo{person}{Ying Li}, \bibinfo{person}{Qianxiang Wang}, {and} \bibinfo{person}{Tao Xie}.} \bibinfo{year}{2024}\natexlab{}.
\newblock \showarticletitle{CoderEval: A Benchmark of Pragmatic Code Generation with Generative Pre-trained Models}. In \bibinfo{booktitle}{\emph{Proceedings of the 46th IEEE/ACM International Conference on Software Engineering}}. \bibinfo{pages}{1--12}.
\newblock


\bibitem[Zhai et~al\mbox{.}(2023)]%
        {zhai2023sigmoid}
\bibfield{author}{\bibinfo{person}{Xiaohua Zhai}, \bibinfo{person}{Basil Mustafa}, \bibinfo{person}{Alexander Kolesnikov}, {and} \bibinfo{person}{Lucas Beyer}.} \bibinfo{year}{2023}\natexlab{}.
\newblock \showarticletitle{Sigmoid Loss for Language Image Pre-Training}. In \bibinfo{booktitle}{\emph{Proceedings of the IEEE/CVF International Conference on Computer Vision}}. \bibinfo{pages}{11975--11986}.
\newblock


\bibitem[Zhang et~al\mbox{.}(2023)]%
        {zhang2023self}
\bibfield{author}{\bibinfo{person}{Kechi Zhang}, \bibinfo{person}{Zhuo Li}, \bibinfo{person}{Jia Li}, \bibinfo{person}{Ge Li}, {and} \bibinfo{person}{Zhi Jin}.} \bibinfo{year}{2023}\natexlab{}.
\newblock \showarticletitle{Self-Edit: Fault-Aware Code Editor for Code Generation}. In \bibinfo{booktitle}{\emph{Proceedings of the 61st Annual Meeting of the Association for Computational Linguistics (Volume 1: Long Papers}}. \bibinfo{publisher}{Association for Computational Linguistics}, \bibinfo{pages}{769--787}.
\newblock


\bibitem[Zhang et~al\mbox{.}(2024)]%
        {zhang2024autocoderover}
\bibfield{author}{\bibinfo{person}{Yuntong Zhang}, \bibinfo{person}{Haifeng Ruan}, \bibinfo{person}{Zhiyu Fan}, {and} \bibinfo{person}{Abhik Roychoudhury}.} \bibinfo{year}{2024}\natexlab{}.
\newblock \showarticletitle{AutoCodeRover: Autonomous Program Improvement}. In \bibinfo{booktitle}{\emph{Proceedings of the 33rd ACM SIGSOFT International Symposium on Software Testing and Analysis}}. \bibinfo{pages}{1592–1604}.
\newblock


\bibitem[Zhang et~al\mbox{.}(2022)]%
        {zhang2022falsifai}
\bibfield{author}{\bibinfo{person}{Zhenya Zhang}, \bibinfo{person}{Deyun Lyu}, \bibinfo{person}{Paolo Arcaini}, \bibinfo{person}{Lei Ma}, \bibinfo{person}{Ichiro Hasuo}, {and} \bibinfo{person}{Jianjun Zhao}.} \bibinfo{year}{2022}\natexlab{}.
\newblock \showarticletitle{FalsifAI: Falsification of AI-Enabled Hybrid Control Systems Guided by Time-Aware Coverage Criteria}.
\newblock \bibinfo{journal}{\emph{IEEE Transactions on Software Engineering}} \bibinfo{volume}{49}, \bibinfo{number}{4} (\bibinfo{year}{2022}), \bibinfo{pages}{1842--1859}.
\newblock


\bibitem[Zhao et~al\mbox{.}(2020)]%
        {zhao2020sim}
\bibfield{author}{\bibinfo{person}{Wenshuai Zhao}, \bibinfo{person}{Jorge~Pe{\~n}a Queralta}, {and} \bibinfo{person}{Tomi Westerlund}.} \bibinfo{year}{2020}\natexlab{}.
\newblock \showarticletitle{Sim-to-Real Transfer in Deep Reinforcement Learning for Robotics: a Survey}. In \bibinfo{booktitle}{\emph{2020 IEEE Symposium Series on Computational Intelligence (SSCI)}}. IEEE, \bibinfo{pages}{737--744}.
\newblock


\bibitem[Zhou et~al\mbox{.}(2023)]%
        {zhou2023specification}
\bibfield{author}{\bibinfo{person}{Yuan Zhou}, \bibinfo{person}{Yang Sun}, \bibinfo{person}{Yun Tang}, {et~al\mbox{.}}} \bibinfo{year}{2023}\natexlab{}.
\newblock \showarticletitle{Specification-Based Autonomous Driving System Testing}.
\newblock \bibinfo{journal}{\emph{IEEE Transactions on Software Engineering}} \bibinfo{volume}{49}, \bibinfo{number}{6} (\bibinfo{year}{2023}), \bibinfo{pages}{3391--3410}.
\newblock


\bibitem[Zhou et~al\mbox{.}(2024)]%
        {zhou2024isr}
\bibfield{author}{\bibinfo{person}{Zhehua Zhou}, \bibinfo{person}{Jiayang Song}, \bibinfo{person}{Kunpeng Yao}, \bibinfo{person}{Zhan Shu}, {and} \bibinfo{person}{Lei Ma}.} \bibinfo{year}{2024}\natexlab{}.
\newblock \showarticletitle{ISR-LLM: Iterative Self-Refined Large Language Model for Long-Horizon Sequential Task Planning}. In \bibinfo{booktitle}{\emph{2024 IEEE International Conference on Robotics and Automation (ICRA)}}. IEEE, \bibinfo{pages}{2081--2088}.
\newblock


\bibitem[Zhu et~al\mbox{.}(2024)]%
        {zhu2023promptbench}
\bibfield{author}{\bibinfo{person}{Kaijie Zhu}, \bibinfo{person}{Jindong Wang}, \bibinfo{person}{Jiaheng Zhou}, {et~al\mbox{.}}} \bibinfo{year}{2024}\natexlab{}.
\newblock \showarticletitle{PromptRobust: Towards Evaluating the Robustness of Large Language Models on Adversarial Prompts}. In \bibinfo{booktitle}{\emph{Proceedings of the 1st ACM Workshop on Large AI Systems and Models with Privacy and Safety Analysis}} (Salt Lake City, UT, USA) \emph{(\bibinfo{series}{LAMPS '24})}. \bibinfo{pages}{57–68}.
\newblock


\end{thebibliography}

\end{document}